\documentclass[a4paper,fleqn,usenatbib]{mnras}

\usepackage{newtxtext,newtxmath}
\usepackage[T1]{fontenc}
\usepackage{ae,aecompl}

\usepackage{graphicx,multirow}	% Including figure files
\usepackage{amsmath}	% Advanced maths commands
\usepackage{amssymb}	% Extra maths symbols
\usepackage{threeparttable}
\usepackage{hyperref}
\usepackage{rotating}
\newcommand{\RN}[1]{%
  \textup{\uppercase\expandafter{\romannumeral#1}}%
}
\title[Short title, max. 45 characters]{The SAMI Galaxy Survey: energy sources of the turbulent velocity dispersion in spatially-resolved local star-forming galaxies}

\author[Luwenjia Zhou et al.]{
Luwenjia Zhou$^{1, 2, 3}$\thanks{E-mail: wenjia@smail.nju.edu.cn},
Christoph Federrath$^{2}$,
Tiantian Yuan$^{2}$,
Fuyan Bian$^{2, 7}$,
\newauthor
Anne M. Medling$^{2, 4, 5}$,
Yong Shi$^{2,3}$,
Joss Bland-Hawthorn$^{8}$,
Julia J. Bryant$^{6, 8, 10}$,
\newauthor
S. Brough$^{13}$,
Barbara Catinella$^{14}$,
Scott M. Croom$^{8,10}$,
Michael Goodwin$^{6}$,
\newauthor
Gregory Goldstein$^{11}$,
Andrew W. Green$^{6}$,
Iraklis S. Konstantopoulos$^{6, 12}$,
\newauthor
J.S. Lawrence$^{6}$,
Matt S. Owers$^{6, 11}$,
Samuel N. Richards$^{8, 9, 10}$,
S. F. Sanchez$^{15}$
\\
$^{1}$School of Astronomy and Space Science, Nanjing University, Nanjing 210093, China\\
$^{2}$Research School of Astronomy \& Astrophysics, Australian National University, Canberra, ACT 2611, Australia\\
$^{3}$Key Laboratory of Modern Astronomy and Astrophysics (Nanjing University), Ministry of Education, Nanjing 210093, China\\
$^{4}$Cahill centre for Astronomy and Astrophysics, California Institute of Technology, MS 249-17, Pasadena, CA 91125, USA\\
$^{5}$Hubble Fellow\\
$^{6}$Australian Astronomical Observatory, PO Box 915, North Ryde NSW 1670, Australia\\
$^{7}$Stromlo Fellow\\
$^{8}$Sydney Institute for Astronomy (SIfA), School of Physics, The University of Sydney, NSW 2006, Australia\\
$^{9}$ Australian Astronomical Observatory, 105 Delhi Rd, North Ryde, NSW 2113, Australia \\
$^{10}$ ARC Centre of Excellence for All-sky Astrophysics (CAASTRO)\\
$^{11}$Department of Physics and Astronomy, Macquarie University, NSW 2109, Australia \\
$^{12}$Atlassian 341 George St Sydney, NSW 2000\\
$^{13}$School of Physics University of New South Wales NSW 2052 Australia\\
$^{14}$ICRAR, M468, The University of Western Australia, 35 Stirling Highway, Crawley, WA 6009, Australia\\
$^{15}$Instituto de Astronom\'\i a, Universidad Nacional Auton\'oma de M\'exico, A.P. 70-264, 04510 M\'exico, D.F.,  Mexico\\
}

\date{Accepted XXX. Received YYY; in original form ZZZ}

\pubyear{2016}

\begin{document}
\label{firstpage}
\pagerange{\pageref{firstpage}--\pageref{lastpage}}
\maketitle

\begin{abstract}

We investigate the energy sources of random turbulent motions of ionised gas from H$\alpha$ emission in eight local star-forming galaxies from the Sydney-AAO Multi-object Integral field spectrograph (SAMI) Galaxy Survey. These galaxies satisfy   strict pure star-forming selection criteria to avoid contamination from active galactic nuclei (AGN) or strong shocks/outflows.  Using the relatively high spatial and spectral resolution of SAMI, we find that -- on sub-kpc scales  our galaxies display a flat  distribution of ionised gas velocity dispersion as a function of star formation rate (SFR) surface density. A major fraction of our SAMI galaxies shows higher velocity dispersion than predictions  by feedback-driven models, especially at the low SFR surface density end.   
 Our results suggest that additional sources beyond star formation feedback contribute to driving random motions of the interstellar medium (ISM) in star-forming galaxies. 
We  speculate that gravity, galactic shear, and/or magnetorotational instability (MRI) may be additional driving sources of turbulence in these galaxies.

\end{abstract}


%%%%%%%%%%%%%%%%%%%%%%%%%%%%%%%%%%%%%%%%%%%%%%%%%%

%%%%%%%%%%%%%%%%% BODY OF PAPER %%%%%%%%%%%%%%%%%%

\section{Introduction}\label{sec:intro}

The kinematics, structure and star formation activity of a galaxy depend on a combination of complex physical processes  such as gravity, turbulence, magnetic fields, radiation, heating/cooling, feedback, accretion,  operating both interior to and exterior to the galaxy. The relative importance of these processes is expected to depend on cosmic evolution and galactic environment.  Galaxies at different cosmic epochs show quite distinct properties. Compared to their high-redshift counterparts at similar stellar masses, local star-forming galaxies are larger, and have relatively  lower gas fractions and lower SFRs \citep{Leroy05, Daddi10, tacconi10, Madau14}. They are also less likely to experience violent events such as major mergers and gas accretion \citep{Baugh96, Genzel08,Robotham14}. 

Many theoretical and observational studies suggest that gas in higher redshift galaxies has larger random motions compared to gas in low-redshift galaxies \citep{Nesvadba06, Lehnert09, Lehnert13, Forster09, Wisnioski15}.  These random, turbulent motions may play a crucial role in regulating the formation of stars  \citep{Green10, Federrath12, Padoan14}. However, the origin and energy source of the turbulence remains poorly understood. External mechanisms like gas accretion from the intergalactic medium and  minor mergers \citep{Glazebrook13}, and internal mechanisms such as star formation feedback (stellar winds, supernovae), cloud-cloud collisions in the disc \citep{Tasker09}, the release of gravitational energy via accretion of cold gas streams from the halo or the inspiral of clumps \citep{Klessen1011}, galactic shear from the differential rotation in disc galaxies \citep{Krumholz15},  spiral-arms shocks in spiral galaxies, magnetorotational instability (MRI)  \citep{Tamburro09} and  others can poteintially drive such turbulence \citep{MacLow04, Elmegreen09, Federrath16b, Federrath17}. 

Several studies have been carried out to explore the energetic drivers of the turbulence in both high and low redshift disc galaxies, e.g., \citet{Elmegreen04, Scalo04, Tamburro09, Gritschneder09}. \citet{Green14} found  that the gas velocity dispersion increases with SFR in star-forming galaxies both locally and at high redshift.
Based on observations and analytic considerations, \citet{Lehnert09, Lehnert13}  speculated that there is a  relation between velocity dispersion and SFR surface density ($\Sigma_{\rm SFR}$) in active star-forming galaxies at \mbox{$z\sim1$--$3$}, and that it is  the intense star formation that supports the high velocity dispersion and thus balances the gravitational pressure.
In contrast, \citet{Genzel11} found that the velocity dispersion correlates only weakly with $\Sigma_{\rm SFR}$ in their study of  giant star-forming clumps in five galaxies at {\it z} $\sim$ 2 together with other rotation-dominant star-forming galaxies, lensed galaxies, and dispersion-dominated galaxies at {\it z} $\sim$ 2. They suggest that a large-scale release of gravitational energy could induce the global large random motions in  high-redshift galaxies, and that local star formation feedback triggering outflows and stirring up the interstellar medium (ISM)  drives the local variation of turbulent, random motions.

Spatially resolved information is vital to understand the details of the physical processes that drive different interactions  within galaxies. The three-dimensional (3D) spectra of galaxies uncover the distribution of the physical properties and give clues to how the internal physical processes shape the galaxies by connecting the spectral information with its position in the galaxy.  Integral field spectroscopy (IFS)  enables us to obtain this crucial spatial information. More importantly, IFS gives us both spectral and kinematic information; i.e., the intensity-weighted gas velocity and velocity dispersion along the line of sight. 
Taking advantage of this, IFS surveys such as  the Calar Alto Legacy Integral Field Area Survey \citep[CALIFA;][]{Sanchez12}, the SAMI Survey \citep{Croom12, Bryant15}, and the Mapping Nearby Galaxies at Apache Point Observatory (MaNGA) Survey \citep{Bundy15} have made significant progress in this area. SAMI has a higher instrumental resolution $\sigma$ = 29 km s$^{-1}$ at 6250 -- 7350 $\AA$ \citep{Sharp15}  than the  other  IFS surveys mentioned above, which have  instrumental resolution > 80 km s$^{-1}$ at a similar wavelength range \citep{Sanchez15}. In this work we investigate the properties of the ISM in star-forming galaxies using data from the SAMI Galaxy Survey. 
We measure maps of $\Sigma_{\rm SFR}$ and gas  velocity dispersion, and use these maps to derive the relation between $\Sigma_{\rm SFR}$ and gas velocity dispersion. We further include  $\Sigma_{\rm SFR}$ and velocity dispersion data from the literature, and determine the dependence of velocity dispersion on redshift, up to {\it z} $\sim$ 3.

Section \ref{sec:data} presents  the sample selection criteria and data reduction including the signal-to-noise ratio criteria, the data source and their reduction strategy, and the estimation of the magnitude of beam smearing effect. In Section \ref{sec:results}, our results and  comparison with high-redshift and H$\alpha$ luminous local star-forming galaxies are presented. We further discuss the main source(s) of the turbulence in star-forming galaxies in Section \ref{sec:discussion}. Section \ref{sec:conclusion} summarzies  the conclusions of this work.
A standard cosmology of  $H_0$ = 70 km s$^{-1}$ Mpc$^{-1}$, $\rm \Omega_m$ = 0.3, $\Omega_{\Lambda}$ = 0.7 is assumed throughout.

\section{Sample and Data Analysis}
\label{sec:data}

 %---------------------------------------- table 2: LZIFU ------------------------------------------
\begin{table}
\centering
%\scriptsize
\caption{\label{tab:LZIFU} Red and blue data cubes from LZIFU.}
\begin{threeparttable}
\begin{tabular}{lccccc}
\hline
 data cube &$\lambda$\tnote{$\dagger$}        &R\tnote{$\ddagger$}  &$\sigma$\tnote{$\ast$} \\
\hline
Blue		 &3700 -- 5700 $\AA$   	&1730 	&74 km s$^{-1}$\\
Red 	     &6250 -- 7350 $\AA$  	&4500 	&29 km s$^{-1}$\\
\hline
\end{tabular}
\begin{tablenotes}
        \item[$\dagger$]Wavelength range.
        \item[$\ddagger$]Spectral resolution. Full width half  maximum (FWHM) = {\it c}/R.
        \item[$\ast$]Velocity resolution according to  spectral resolution. %\quad $\sigma$ =  $\frac{\rm FWHM}{\rm 2\sqrt{2ln2}}$ = $\frac{\rm FWHM}{\rm 2.35}$  =   $\frac{{\it c} /\rm  R}{\rm 2.35}$.  
      \end{tablenotes}
      \end{threeparttable}
\end{table}

%---------------------------------------- end of table 2 ------------------------------------------

%----------------------------------------  table 1: 9 sf galaxies ------------------------------------------
\begin{table*}
\centering
\footnotesize
\caption{\label{tab:properties} Properties of  the eight star-forming galaxies in our final sample of star-forming SAMI galaxies.}
\begin{threeparttable}
\begin{tabular}{rrrrrrlrrrr}
\hline
  CATID     	&RA   		   &DEC          		&redshift	& stellar mass\tnote{1}	&\multicolumn{2}{c}{radius\tnote{2}}     &ellip\tnote{3} &{\it i }\tnote{4} &$\sigma_{\rm gas}$\tnote{5} &$\Sigma_{\rm SFR}$\tnote{6}\\
                &[hh: mm: ss]  &[dd: mm: ss]     	&   		&[M$_{\odot}$]			& [$\arcsec$]   		     &[kpc]                 &		 & [\textdegree]			&	[km s$^{-1}$]	  &[$\rm M_{\odot}$ yr$^{-1}$ kpc$^{-2}$]\\
\hline
79635	&14 50 03.3  & 00 05 51.0 	&0.040 	&2.9 $\times 10^{10}$  	&9.13	&7.8 	&0.40	&55.0	&28 $\pm$ 4\:\:		&0.019 $\pm$ 0.009\\%2.70 \\
376001 	&08 46 31.3	 & 01 29 00.1	&0.051 	&1.8 $\times 10^{10}$ 	&2.41	&2.7	&0.07	&22.4	&31 $\pm$ 9\:\:		&0.022 $\pm$ 0.007\\%2.94\\
388603 	&09 23 08.1	 & 02 29 09.9 	&0.017 	&6.3 $\times 10^{9}$\:\:	&14.3	&5.2	&0.12	&28.6	&24 $\pm$ 4\:\:		&0.009 $\pm$ 0.003\\%2.79 \\
485885 	&14 31 01.9	 &-01 43 02.0 	&0.055 	&1.8 $\times 10^{10}$ 	&5.04	&6.0	&0.16	&33.6	&24 $\pm$ 4\:\:		&0.014 $\pm$ 0.005\\%3.87 \\
504882 	&14 30 15.3  &-01 55 56.2  	&0.054 	&1.3 $\times 10^{10}$ 	&3.80	&4.4	&0.19	&37.0	&20 $\pm$ 2\:\:		&0.010 $\pm$ 0.003\\%2.75 \\
508421 	&14 27 57.4	 &-01 37 52.3 	&0.055 	&2.5 $\times 10^{10}$ 	&3.74	&4.5	&0.26	&43.0	&87 $\pm$ 44		&0.076 $\pm$ 0.016\\%2.59 \\
599582 	&08 48 45.6  & 00 17 29.5	&0.053 	&6.2 $\times 10^{10}$ 	&9.60	&11	&0.32	&48.6	&26 $\pm$ 5\:\:		&0.020 $\pm$ 0.009\\%2.14 \\
618152 	&14 18 05.5	 & 00 13 38.6 	&0.053 	&1.0 $\times 10^{10}$ 	&3.56	&4.1	&0.29	&46.1	&24 $\pm$ 3\:\:		&0.023 $\pm$ 0.010\\%4.21 \\
%618220 	&14 18 57.4  & 00 21 56.3 	&0.053 	&3.9 $\times 10^{10}$ 	&4.90	&5.64	&0.17	&34.8	&47.7 $\pm$ 9.4	&5.1 $\pm$ 1.3\\%1.24 \\
\hline
\end{tabular}
\begin{tablenotes}
        \item[1] Stellar masses are from the GAMA survey \citep{Taylor11}.      
        \item[2] Effective radius, i.e., half light radius, also from the GAMA survey \citep{Kelvin12}.
        \item[3] Ellipticity is from the GAMA survey (\href{url}{http://www.gama-survey.org/dr2/tools/sov.php}). We use the GAL\_ELLIP\_R to get the R-band axis ratio.  The relation between minor-to-major axis ratio and ellipticity is: {\it b/a} = 1 - ellipticity.
        \item[4]  Inclination angle. The calculation is based on classical Hubble formula:  cos$^{2}i = ((b/a)^2 - q_0^2)/(1 - q_0^2))^{1/2}$, where {\it b/a} is the minor-to-major axis ratio, {\it i}  is the inclination angle and {\it q$_{0}$} = 0.2 ({\it i} = 90\textdegree \: for {\it b/a} < $q_{0}$). 
        \item[5]  Flux weighted global gas velocity dispersion. Only the pixels with $\sigma_{\rm gas}$ > 2 $v_{\rm grad}$ are considered (see more in Section \ref{sec:beamsmearing}).
        \item[6]  Flux weighted SFR surface density. Only the pixels with $\sigma_{\rm gas}$ > 2 $v_{\rm grad}$ are considered (see more in Section \ref{sec:beamsmearing}).
      \end{tablenotes}
      \end{threeparttable}
\end{table*}  

%---------------------------------------- end of table 1 ------------------------------------------

\subsection{Sample Selection}
%-----------------------figure 1 : SFR, Vgas, VDISPgas, VDISPstar, STELLAR MASS images ------------------------------------------
\begin{figure*}
\centerline{
\setlength{\tabcolsep}{2pt}
\LARGE
\begin{tabular}{ccccc}
79635&376001&388603&485885\\
 \includegraphics[width=0.22\linewidth]{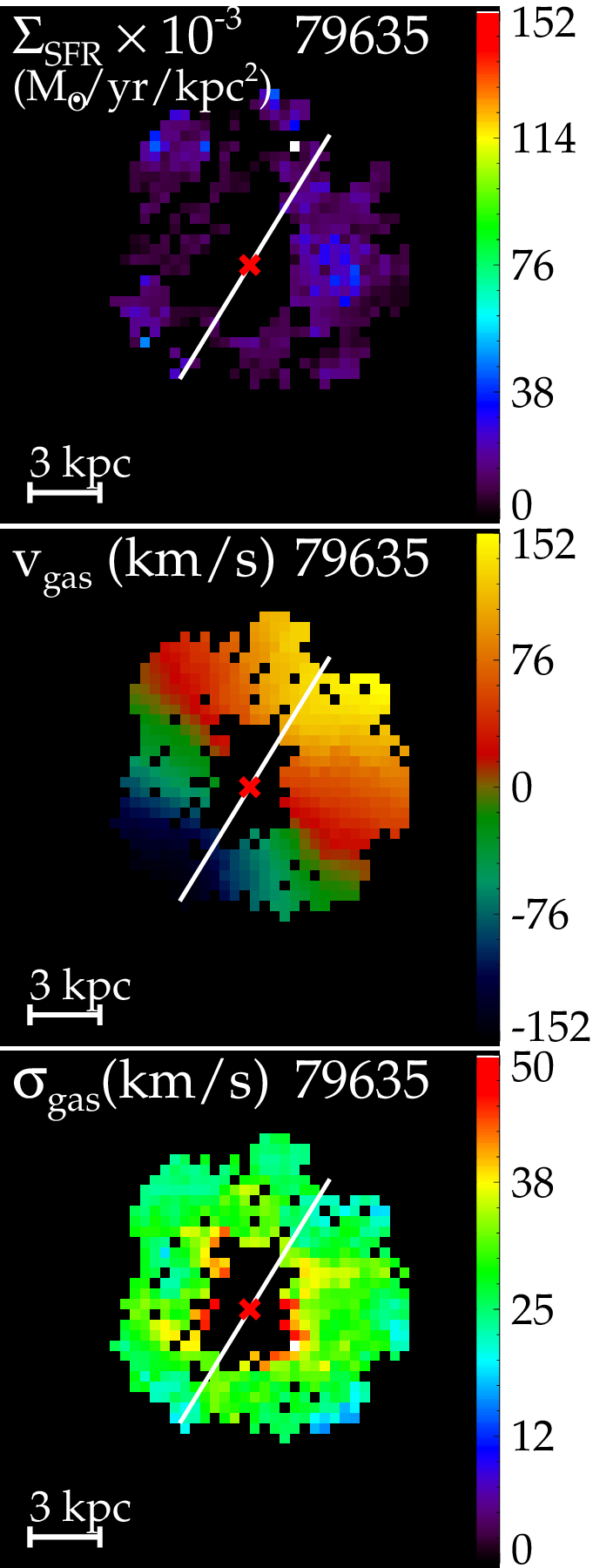} &
\includegraphics[width=0.22\linewidth]{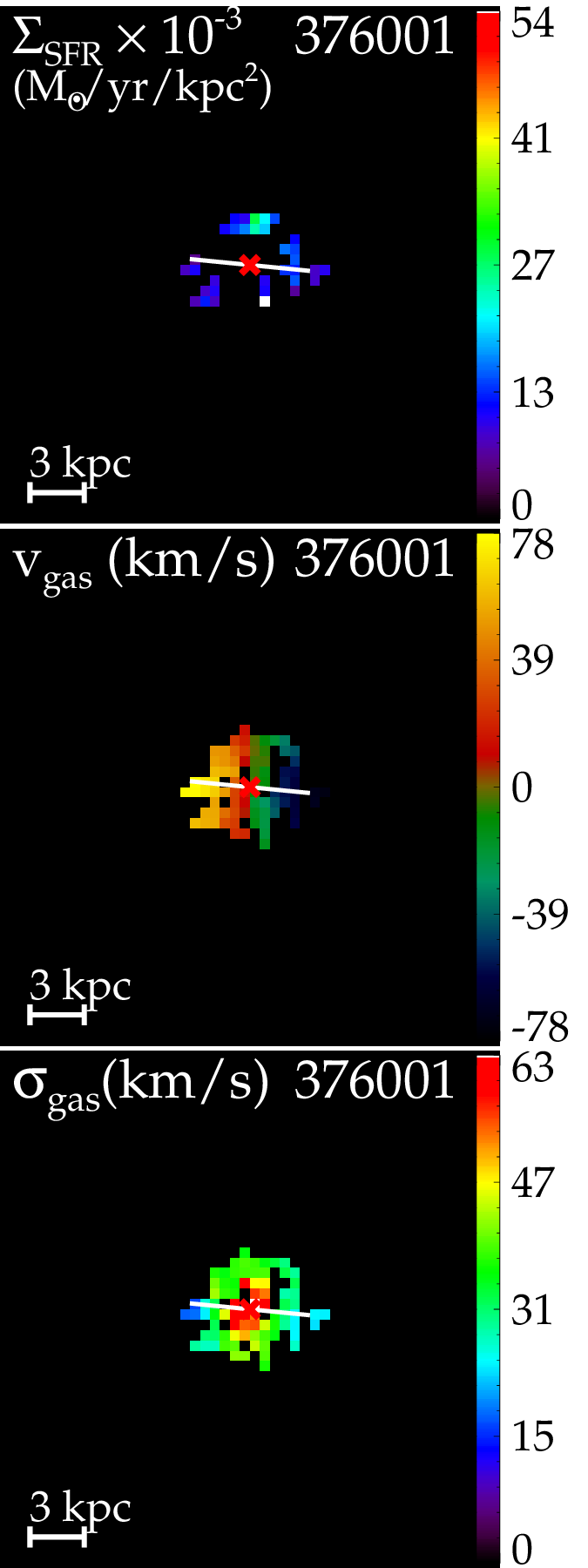} &
\includegraphics[width=0.22\linewidth]{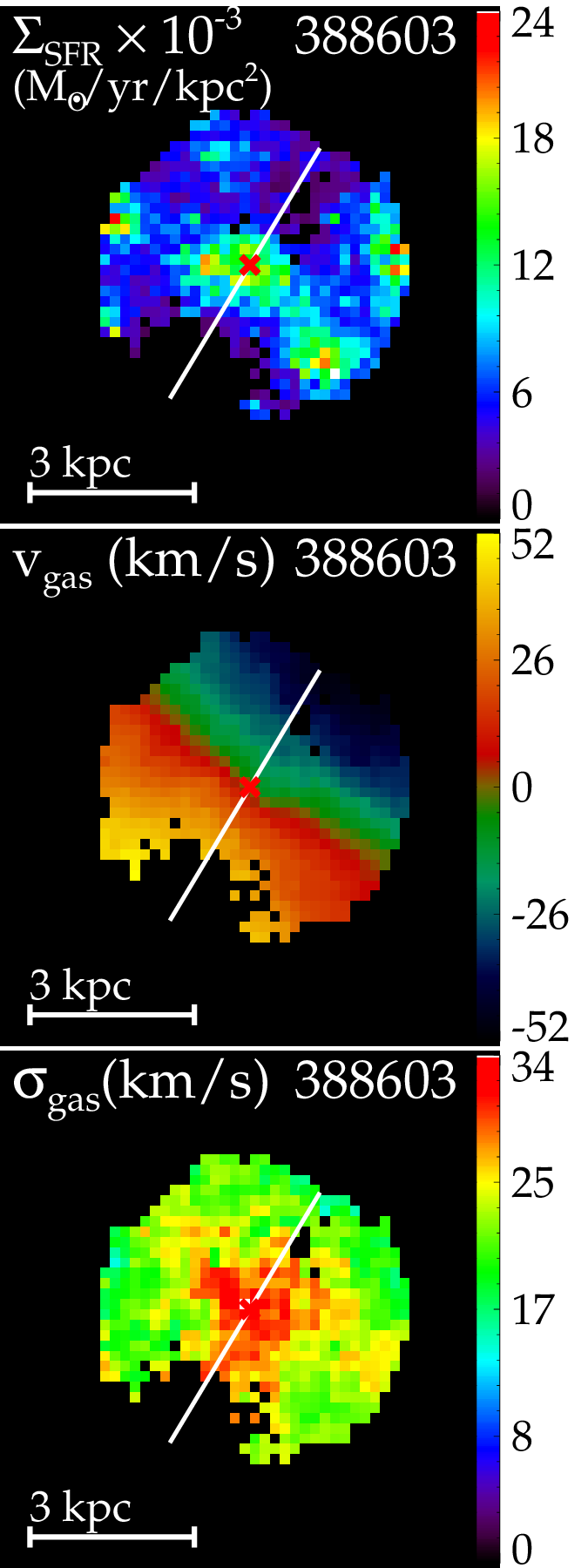} &
\includegraphics[width=0.22\linewidth]{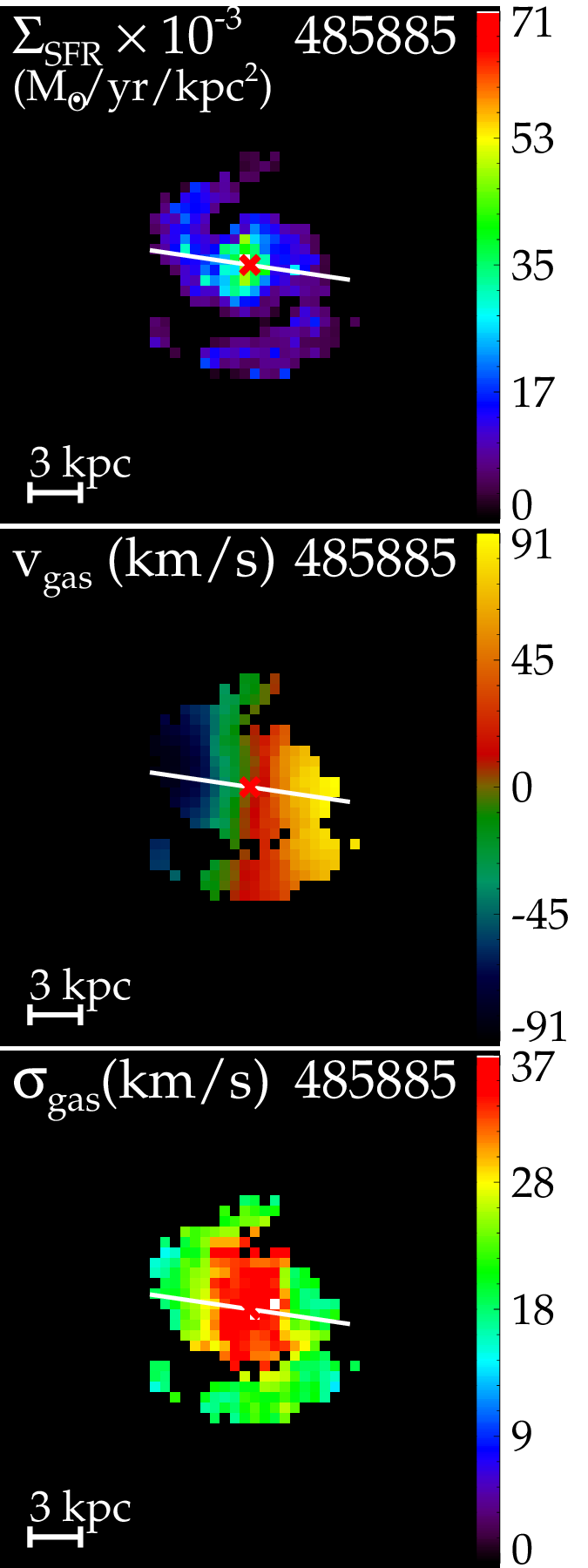} \\
504882&508421&599582&618152\\
\includegraphics[width=0.22\linewidth]{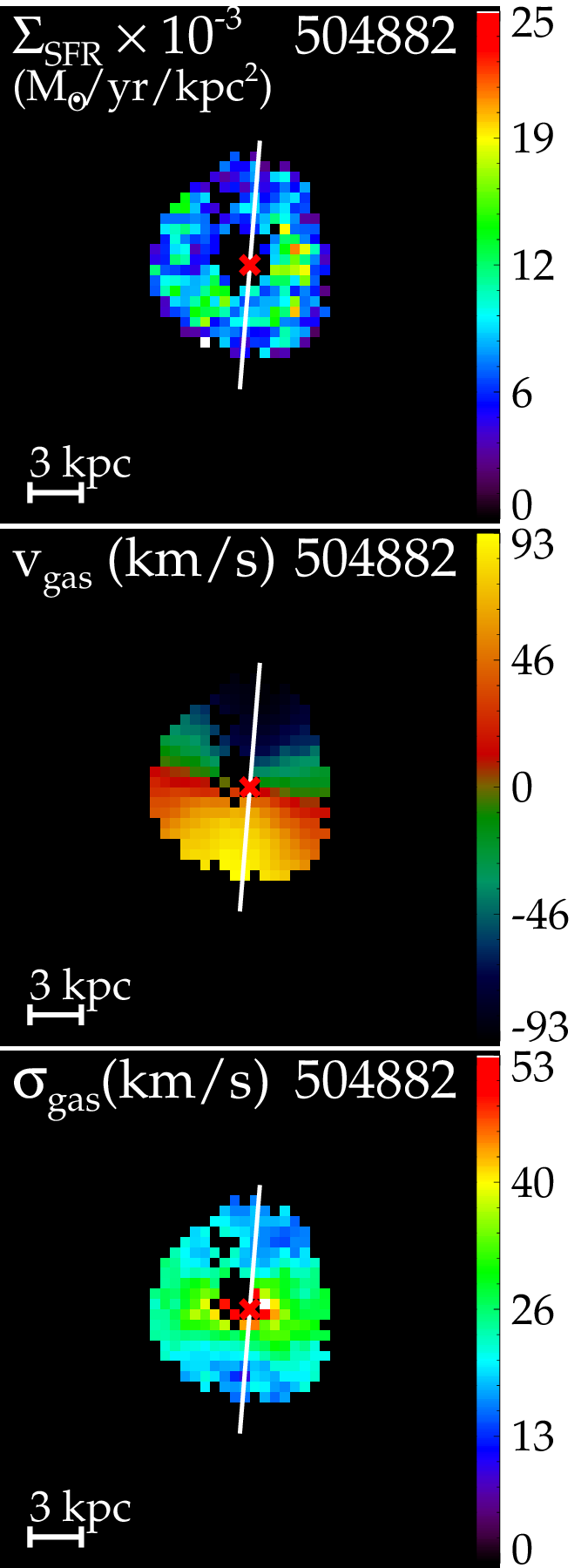} &
\includegraphics[width=0.22\linewidth]{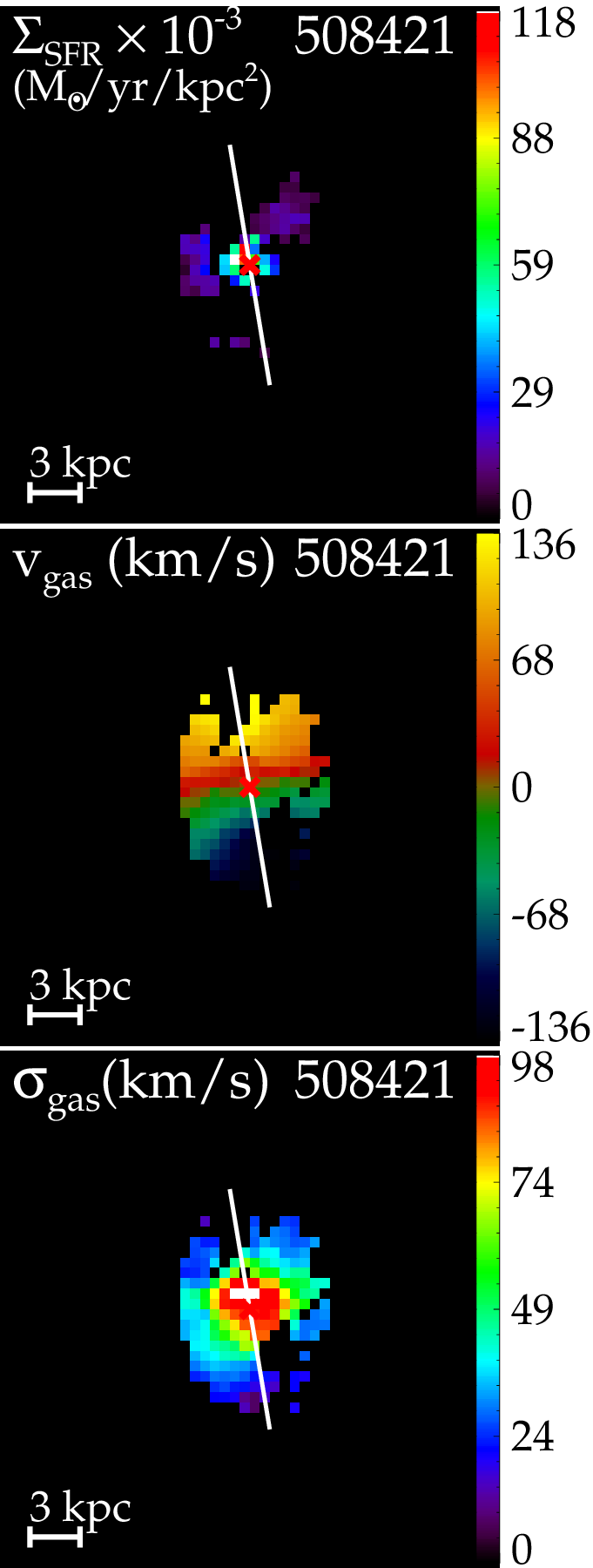} &
\includegraphics[width=0.22\linewidth]{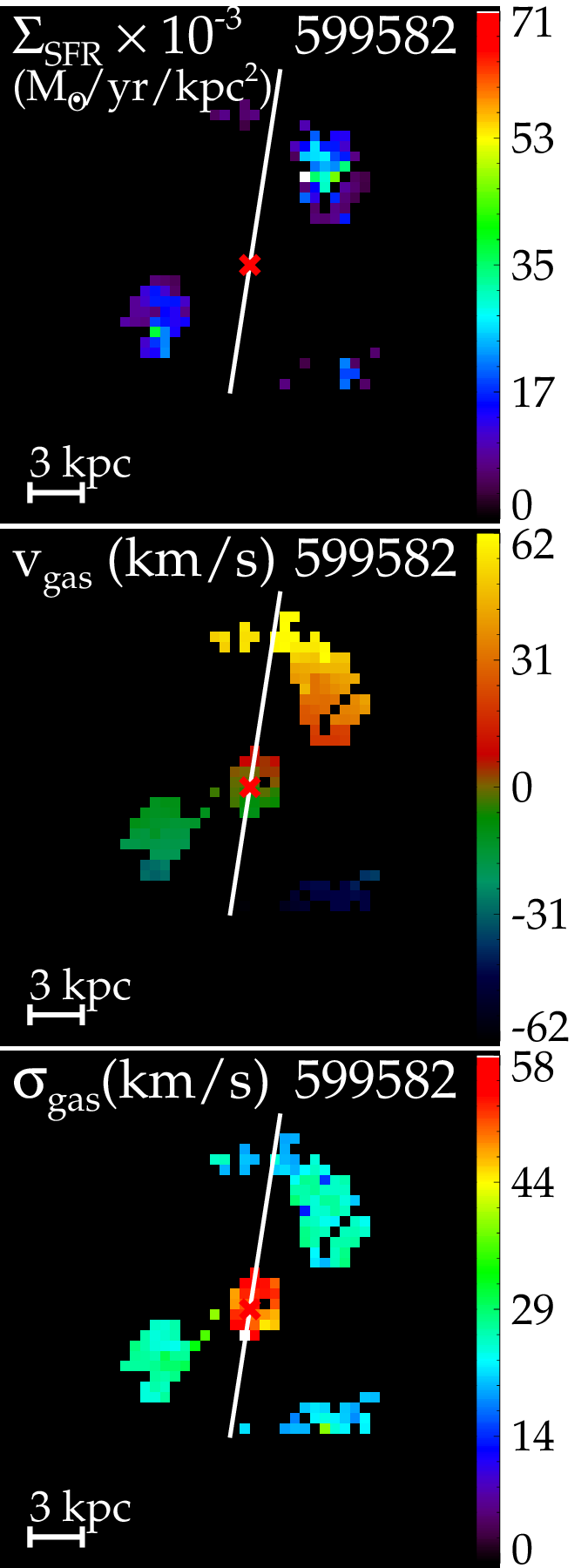} &
\includegraphics[width=0.22\linewidth]{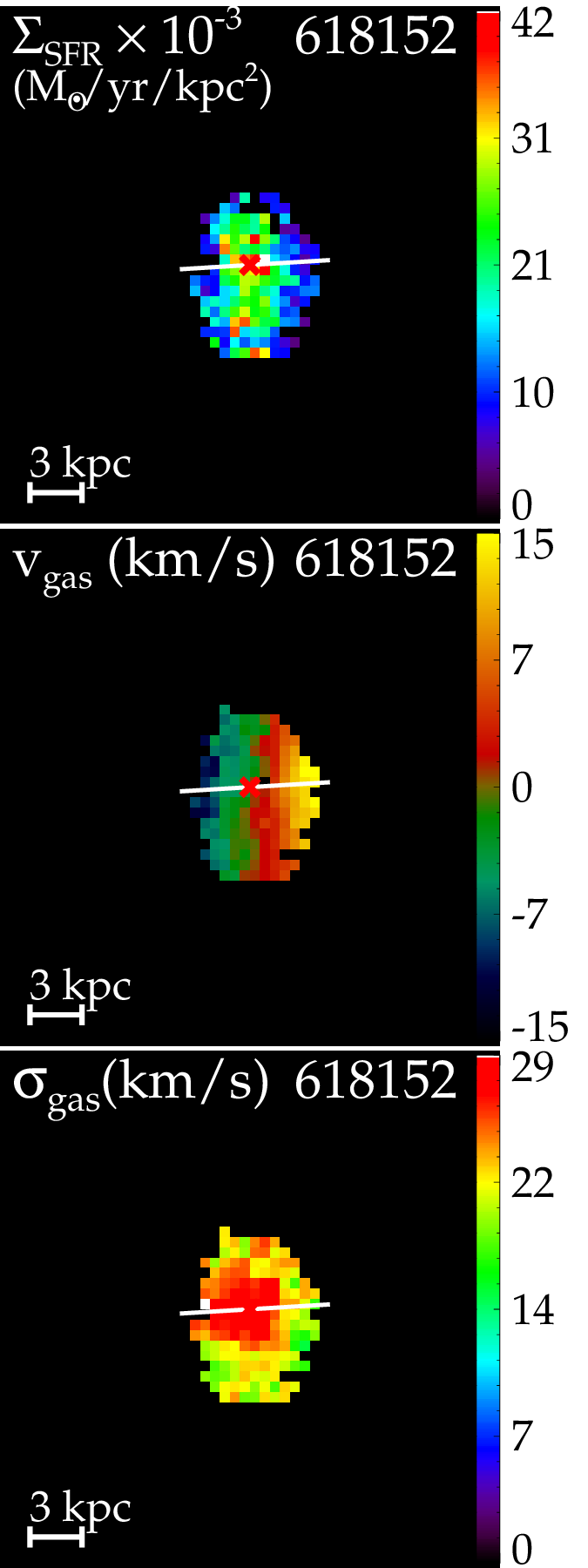} \\
\end{tabular}}
\caption{Images of different components of each galaxy. Top: SFR surface density (extinction corrected) in M$_{\odot}$ yr$^{-1}$ kpc$^{-2}$. Middle: ionised gas velocity in km s$^{-1}$. Bottom: ionised gas velocity dispersion in km s$^{-1}$. In all images, we only display spaxels with S/N  > 34  (see Section \ref{sec:snr} for the choice of S/N). The red crosses are the centres of the galaxies. The white solid lines are the major axes (see Section \ref{sec:majoraxis}).}
\label{fig:images}
\end{figure*}
  
%---------------------------------------- end of figure 1 ------------------------------------------

\subsubsection{The SAMI Galaxy Survey}

We use the SAMI Galaxy Survey \citep{Croom12} to select local star-forming galaxies.  
The SAMI Survey will observe in total $\sim$ 3400 galaxies. It covers a broad range of galaxies in stellar mass and environment. The sample targets redshifts 0.004 -- 0.095, Petrosian magnitudes r$_{\rm pet}$ < 19.4 \footnote{Extinction-corrected SDSS DR7 Petrosian mag.}, stellar masses 10$^{7}$ -- 10$^{12}$ M$_{\odot}$ and environments from isolated field galaxies through group galaxies to cluster galaxies \citep{Bryant15, Owers17}.

SAMI is mounted at the prime focus on the Anglo-Australian Telescope. It  has a 1 degree diameter field of view and uses 13 fused fibre bundles \citep[Hexabundles,][]{BlandHawthorn11, Bryant14} with a high (75\%) fill factor. Each of the bundles contains 61 fibres of 1.6 arcsec diameter which results in  diameter of 15 arcsec in each integral field unit (IFU). The IFUs, together with 26 sky fibres, are inserted into pre-drilled plates using magnetic connectors. SAMI fibres are fed to the double-beam AAOmega spectrograph \citep{Sharp06}. AAOmega provides a range of different resolutions and wavelength ranges. For the SAMI Galaxy Survey,  the 570V grating at 3700 -- 5700 $\AA$ is used to give a resolution of R = 1730 ($\sigma$ = 74 km s$^{-1}$), and the R1000 grating from 6250 -- 7350 $\AA$ for a resolution of R = 4500 ($\sigma$ = 29 km s$^{-1}$) (Table \ref{tab:LZIFU}) \citep{Sharp15}. Therefore, each SAMI object has one blue and one red data cube.  Early reduced datacubes are included in the first public data release \citep[][Green et al. in prep.]{Allen15}.
 All datacubes are constructed on spatial grids where each grid cell has a size of $\rm 0.5\arcsec \times 0.5\arcsec$   corresponding to a physical scale of $\sim$ 0.5 $\times$ 0.5 kpc$^2$ for our sample of local star-forming galaxies ({\it z} $\sim$ 0.05). Note that the spatial resolution of SAMI data is determined by the average seeing in observations ($\sim$ 2.5$\arcsec$), corresponding to a physical scale of 2.5 kpc (z $\sim$ 0.05).

The spectral fitting pipeline of SAMI galaxies, LZIFU \citep{Ho14,Ho16}, is designed to extract  two-dimensional emission line flux maps and kinematic maps to investigate the dynamics of gas in galaxies. LZIFU uses up to three Gaussian profiles to fit emission lines, separating up to three different kinematic components contributing to the emission lines. %For this work, 

Maps of SFR and SFR surface density (Medling et al. in prep.) are also available in the SAMI database.  Briefly, these maps are calculated from extinction-corrected H$\alpha$ flux maps using the calibration in \citet{Kennicutt94a}. Extinction corrections are calculated using the Balmer decrement, and flux is converted to luminosity using distances calculated from the flow-corrected redshifts of the GAMA Survey Catalog \citep{Baldry12}.

\subsubsection{Our sample}
\label{sec:sample}
To select star-forming galaxies from the parent SAMI sample, we use optical emission line diagnostic  diagrams, so-called 'BPT/VO' diagrams \citep{Baldwin81, Veilleux87} with multi-component emission line fits. In the following analysis of the chosen star-forming galaxies, we use single component fits, which are sufficient to describe  star-forming galaxies. In the most conservative way, we select those galaxies with all  detected spaxels lying below the theoretical extreme starburst lines in all the three BPT/VO87 diagrams, i.e., [N$\RN{2}$]$\rm\lambda$6583/H$\alpha$ {\it vs.} [O$\RN{3}$]$\rm\lambda$5007/H$\beta$, [S$\RN{2}$]$\rm\lambda$6717,$\rm\lambda$6731/H$\alpha$ {\it vs.} [O$\RN{1}$]$\rm\lambda$6300/H$\beta$ and [N$\RN{2}$]$\rm\lambda$6583/H$\alpha$ {\it vs.} [O$\RN{3}$]$\rm\lambda$5007/H$\beta$  \citep{Baldwin81, Veilleux87, Kewley01}. Thus we  minimise the contamination of AGNs, outflows, and shock contamination. We note that there are a few spaxels  with elevated non-thermal-ratios suggestive of  shocks in 508421. Because these spaxels are few in numbers and lie below the theoretical extreme starburst lines in all the three BPT/VO diagrams, we still keep this galaxy.

We find that eight out of 756 SAMI galaxies satisfy our complete selection criteria. By the time of this work, 756 of the 3400 SAMI galaxies  have had reduced data cubes available with the required analysis products. We  find 22 star-forming galaxies meeting the criteria, but 11 of them are cluster galaxies which may be influenced by their environments.  We  choose not to discuss cluster galaxies, in order to focus on turbulence induced locally through accretion and/or galaxy-internal processes such as star formation feedback. Among the remaining 11 galaxies,  two lack the information of stellar velocity dispersion \citep[][this will be relevant for a follow-up paper to determine the Toomre Q and that we want to use the same set of eight galaxies for this and the follow-up paper]{Sande17} and another  lacks enough spaxels with high enough signal-to-noise ratio. Table \ref{tab:properties} lists the basic information of the eight galaxies in our final sample. Our eight star-forming galaxies are at redshifts ranging from  0.017 to 0.055, most of them at the high end. Their stellar masses range from 6.3 $\times 10^{9}$ M$_{\odot}$ to 6.2 $\times 10^{10}$ M$_{\odot}$ and the median stellar mass is 2.5 $\times 10^{10}$ M$_{\odot}$, similar to most of the galaxies in SAMI sample \citep{Bryant15}.

\subsection{Gas  Kinematic Information}
\label{sec:data1}
\subsubsection{Definition of the major axis}
\label{sec:majoraxis}
To define the major axis, we first define the centre of each galaxy as the centre of each datacube. We note that this is consistent with  the photometric and kinematic centres of each galaxy in our sample. The major axis of each galaxy is determined based on the velocity field of the galaxy. The centre velocity ($v_{\rm centre}$) of each galaxy is measured by averaging the central four pixels; gas velocity is given as: $v_{\rm gas}$ = $v - v_{\rm centre}$. 

\subsubsection{Velocity and Velocity dispersion}%S/N requirement for gas velocity dispersions
\label{sec:snr}

 Ionised gas velocity ($v_{\rm gas}$) and gas velocity dispersion ($\sigma_{\rm gas}$) are measured from the emission lines by the LZIFU pipeline.
 The  velocity dispersions in the datacubes have removed instrument resolutions, i.e.,  $\rm \sigma_{gas}= (\sigma_{\rm obs}^2-\sigma_{\rm instr}^2)^{1/2}$, where $\sigma_{\rm instr}$  is the instrumental velocity dispersion and $\sigma_{\rm obs}$ is the observed velocity dispersion). 
We emphasise that the error of the line width  can be underestimated because of the limitation of the instrument resolution. Thus we make an estimation of the lower limit of the reliable velocity dispersion. Given that $\sigma_{\rm instr}$ = 29 km $ \rm s^{-1}$ (assuming the instrument resolution for the H$\alpha$ line  is fixed and exactly 29 km s$^{-1}$), if we want to resolve an intrinsic velocity dispersion of 12 km $ \rm s^{-1}$  with signal to noise ratio (S/N) of 5 (i.e., (S/N)$_{\rm true}$ $\equiv \frac{\sigma_{\rm gas}}{d\sigma_{\rm gas}}$ = 5, $\sigma_{\rm gas}$ is the intrinsic velocity dispersion in the datacubes), then we can derive: 
\begin{equation}
\begin{aligned}
 &\rm \sigma^2_{gas} =  \sigma^2_{obs} -  \sigma^2_{instr} ; \nonumber \\
 &\rm d\sigma^2_{gas} =  d\sigma^2_{obs} - d\sigma^2_{instr}\, ,\nonumber \\
 &\rm (d\sigma^2_{instr}=2 d\sigma_{instr}=0\, ,\ \sigma_{instr}=29\; km\; s^{-1}) ; \nonumber \\
 &\rm 2 \sigma_{gas} d\sigma_{gas} = 2 \sigma_{obs} d\sigma_{obs} ; \nonumber \\
 &\rm dividing\: both \: sides \: with \: \sigma_{ gas}^2: \\
 &\rm \frac{d\sigma_{gas}}{\sigma_{gas}}=\frac{d\sigma_{obs}}{\sigma_{obs}} \   \frac{\sigma^2_{obs}}{\sigma^2_{gas}} ; \nonumber \\
 &\rm rearranging \: this  \: equation \: and  \:  substituting  \: \sigma_{\rm obs}^2  \: with  \:  \sigma_{\rm gas}^2+\sigma_{\rm instr}^2: \\
&\rm \frac{\sigma^2_{obs}}{\sigma^2_{gas}}=\rm \frac{\sigma^2_{gas}+\sigma^2_{instr}}{\sigma^2_{gas}}=\frac{\sigma_{obs}}{d\sigma_{obs}}\Bigg/ \frac{\sigma_{gas}}{d\sigma_{gas}} \equiv \frac{(S/N)_{\rm obs}}{(S/N)_{\rm true}}; \nonumber \\
&\rm \frac{\sigma_{obs}}{d\sigma_{obs}} = \frac{\sigma_{gas}}{d\sigma_{gas}}(\frac{\sigma_{instr}^2}{ \sigma_{gas}^2}+1),\  i.e., \nonumber \\
&\rm  (S/N)_{\rm obs} =  (S/N)_{\rm true} (\frac{\sigma_{instr}^2}{ \sigma_{gas}^2}+1) = 5 \times (\frac{29^2}{12^2}+1) = 34.
\end{aligned}
\end{equation}
Then observed emission line S/N needs to be 34, i.e., (S/N)$_{\rm obs}$ $\equiv \frac{\sigma_{\rm obs}}{d\sigma_{\rm obs}}$ = 34. 
This criterion translates to a measured velocity dispersion S/N cut that depends on velocity dispersion:
(S/N)$_{\rm true}$ = (S/N)$_{\rm obs}$ / ($\sigma_{\rm instr}^2 / \sigma_{\rm gas}^2$ + 1).
Therefore, in order to resolve velocity dispersion down to 12 km $ \rm s^{-1}$ (corresponding to the thermal broadening velocity of ionised gas) with S/N > 5, we select only  spaxels with S/N > 34 for velocity dispersion.
Given that the velocity dispersion we are measuring is lower than the spectral resolution, we  do  simple Monte Carlo simulations to see if there is a systematic overestimation. The simulations test (S/N)$_{\rm obs}$ from 10 to 34 and a range of line widths between  5 km s$^{-1}$ and 40 km s$^{-1}$. They confirm the scaling predicted by our very simple analytic model: indeed, in order to measure $\sigma_{\rm gas}$ (with a target intrinsic S/N) one needs a much larger S/N on the observed.  Assuming a velocity sampling of 5 km/s, we obtain an excellent fit to our simple analytic formula above, i.e., a (S/N)$_{\rm obs}$ > 34 is required for 12 km s$^{-1}$  with (S/N)$_{\rm true}$ > 5. However, the absolute S/N requirement eventually depends on the specifics of the instrument, and a separate publication fully dedicated to these Monte Carlo simulations is in preparation.

Maps of gas velocity ($v_{\rm gas}$) and gas velocity dispersion ($\sigma_{\rm gas}$) together with $\Sigma_{\rm SFR}$ are shown in Figure \ref{fig:images}.

\subsubsection{Beam smearing effect}
%\label{sec:v_grad}
\label{sec:beamsmearing}
%----------------------------------------------- figure 2: shear vs. vdisp -----------------------

\begin{figure}
\includegraphics[width=1.\linewidth]{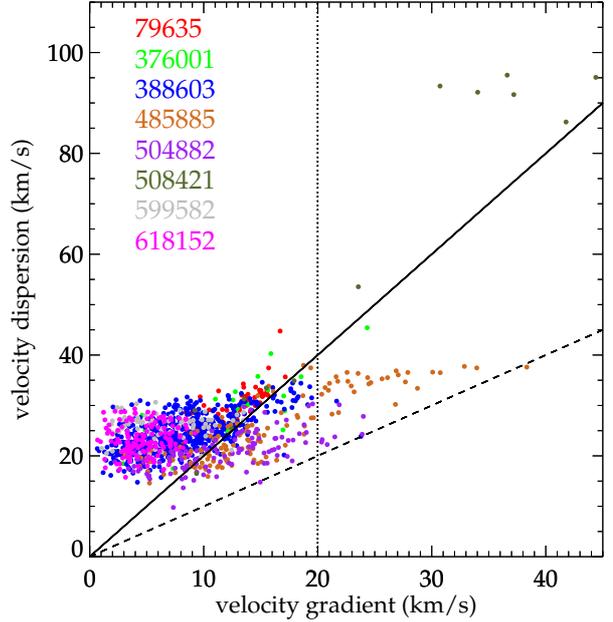} 
\caption{Dependence of gas velocity dispersion ($\sigma_{\rm gas}$) on velocity gradient ($v_{\rm grad}$, Equation \ref{eqn:grad}).  Points in each individual galaxy are labeled by colour. The solid  line refers to $\sigma_{\rm gas}$ = 2 $v_{\rm grad}$ and the dashed line refers to $\sigma_{\rm gas}$ = $v_{\rm grad}$. The  dotted line denotes the velocity gradient at 20 km s$^{-1}$. 
These three different lines represent the three different criteria that we tested to account for beam smearing (Section \ref{sec:beamsmearing}). }
\label{fig:grad_contrast}
\end{figure}

%----------------------------------------------- end of figure 2 -------------------------------------------------------

The measured velocity dispersion may be overestimated in the presence of a velocity gradient within the scale of one beam;  this is what we call beam smearing. The magnitude of beam smearing increases with increasing local velocity gradient. We use the velocity gradient ($ v_{\rm grad}$) across a spaxel \citep{Varidel16}, 
\begin{equation}
 v_{\rm grad} = \sqrt{ (v[x+1,y] - v[x-1,y])^2+ (v[x,y+1] - v[x,y-1])^2}. 
\label{eqn:grad}
\end{equation}
 to estimate the magnitude of the beam smearing ({\it x, y} are the image grid cell indices). We use the upper, lower, left and right neighbours of each  spaxel to measure  $ v_{\rm grad}$. These 5 pixels make up the central cross in a 3 $\times$ 3 binned spaxel, which is roughly equivalent to the size of the seeing limited FWHM  of $\sim$2.5$\arcsec$.  Therefore, $v_{\rm grad}$ is indicative of the beam smearing effect within the size of the beam FWHM. A higher $ v_{\rm grad}$ can indicate a larger beam smearing effect and thus a higher observed velocity dispersion. In order to account for the effect of beam smearing in the following analysis, we remove the spaxels   whose line-width may have been primarily caused by a velocity gradient due to beam smearing.  Note that the boundary spaxels don't have enough necessary neighbouring spaxels to calculate velocity gradient, we choose to also remove boundary spaxels.

In Figure \ref{fig:grad_contrast}, we show the measured velocity dispersion as a function of the velocity gradient for each spaxel and galaxy. 
We tried three different criteria to account for the beam smearing effect. We exclude spaxels with:
\begin{enumerate}
\item $ v_{\rm grad}$ > 0.5 $\sigma_{\rm gas}$ (solid line),
\item $ v_{\rm grad}$ > $\sigma_{\rm gas}$ (dashed line),
\item $ v_{\rm grad}$ > 20 km s$^{-1}$ (dotted line). 
\end{enumerate}    We find that our results do not depend on the particular choice of the beam-smearing cut. All three selection criteria yield consistent results. A flat but elevated distribution of gas velocity dispersion (see more in Section \ref{sec:sfrsigma}) is shown in all of the three cases.
We choose to preserve only the spaxels with $\sigma_{\rm gas}$ > 2 $ v_{\rm grad}$ (those above the black solid line)  in the following analysis. Our method here is similar to the simple analytic calculation in \citet[][equation (1)]{Bassett14}.

\subsection{Spatial resolution}
We compare our sample with  high redshift  surveys and local H$\alpha$ luminous galaxies. The data in \citet{Lehnert13} has FWHM $\sim$0.6$\arcsec$ and pixel scale 0.25$\arcsec$ corresponding to 5 kpc and 2 kpc at {\it z} $\sim$ 2. The seeing limit of our sample is $\sim$ 2.5$\arcsec$ and  the pixel scale is 0.5$\arcsec$, which correspond to 2.5 kpc and 0.5 kpc at {\it z} $\sim$ 0.05.  Genzel et al. (in prep.)  has FWHM  up to 0.2$\arcsec$ corresponding to $\sim$ 1.5 kpc at {\it z} $\sim$ 0.76 -- 2.65.  \citet{Green14} has spatial resolutions of 1 -- 3 kpc. We reach similar resolutions as \citet{Genzel11} and \citet{Green14} and better than \citet{Lehnert13}.
 All the works above either construct models or use simulations to remove the beam smearing effect.

\section{Results}
\label{sec:results}

We investigate the relation between $\Sigma_{\rm SFR}$ and $\sigma_{\rm gas}$  spaxel by spaxel (locally), and  within individual galaxies (globally), to see if star formation drives the velocity dispersion in these local star-forming galaxies.

\subsection{The spatial distribution of $\Sigma_{\rm SFR}$, $v_{\rm gas}$, and $ \sigma_{\rm gas}$ } 

In Figure \ref{fig:images}, we show the maps of $\Sigma_{\rm SFR}$, gas velocity and gas velocity dispersion for each of our eight galaxies. The major axes and galaxy centres are labeled in the images. We see:

a) The $\Sigma_{\rm SFR}$ maps have various distributions. There  are often multiple peaks and rings, and some of the peaks are not at the centre,  indicating local structures such as spiral arms,  star-forming clumps, etc. 

b) All  galaxies show clear velocity gradients indicating rotation. 

c)  All  galaxies show a  gas velocity dispersion peak at the centre. %However, the sharp velocity gradient at the centre of the galaxies may also contribute to the large velocity dispersion at the centre. 
%, which is consistent with the distribution of stellar mass.

d) The distribution of gas velocity dispersion does not always follow the distribution of $\Sigma_{\rm SFR}$  (i.e. the peaks in $\sigma_{\rm gas}$  do not always correlate with those in $\Sigma_{\rm SFR}$). This is self-consistent because regions of intense mechanical or radiative energy injections (e.g., star-forming regions) are over-pressured and compact while over-pressurized gas is over a much larger scale.

\subsection{The $\rm \sigma_{gas}$ -- $\rm \Sigma_{SFR}$ relation in local and high redshift  star-forming galaxies}

\subsubsection{Local (spaxel-by-spaxel) analysis}
\label{sec:sfrsigma}
%--------------------------------------- figure 3-1: lehnert09 overplot -------------------------------------------
\begin{figure*}
\centering
	\includegraphics[width=2\columnwidth]{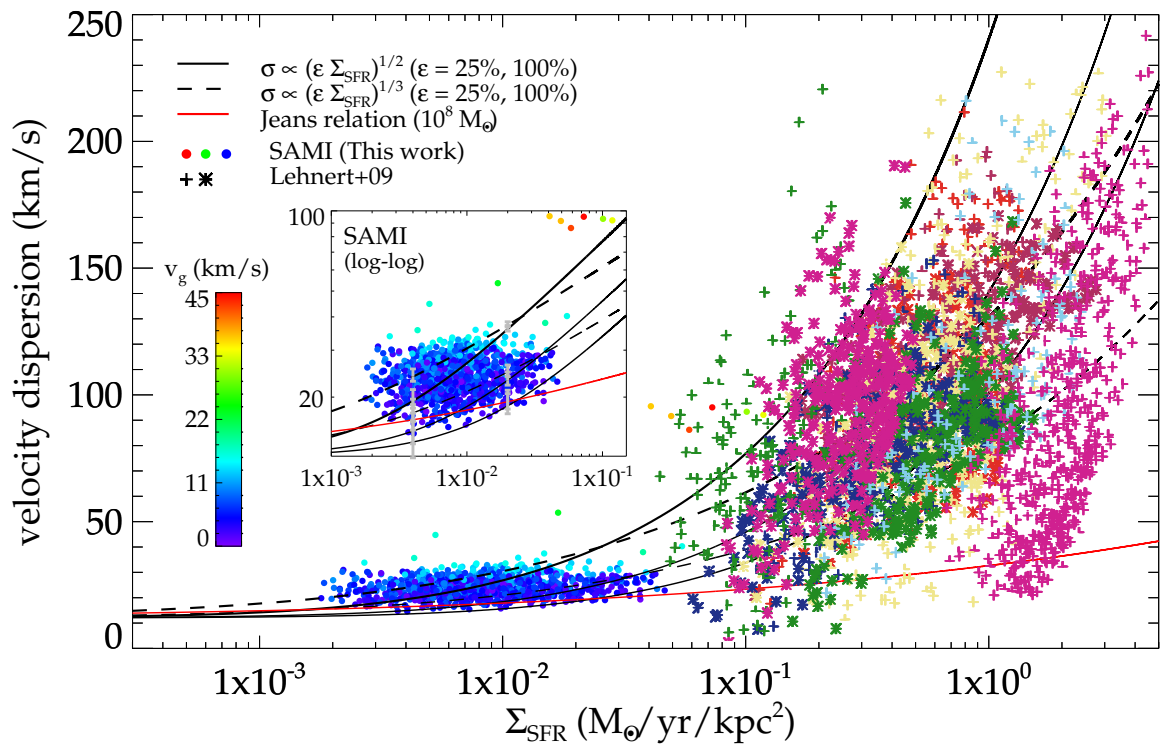}
	\caption{ Spatially resolved dependence of $\sigma_{\rm gas}$ on $\Sigma_{\rm SFR}$.
    We remove pixels with v$_{\rm grad}$ > 0.5 $\sigma_{\rm gas}$ to account for beam smearing effects (see Figure \ref{fig:grad_contrast} and Section \ref{sec:data1}). 
 Our eight SAMI galaxies are compared to \citet[Figure 12 therein]{Lehnert09}.  Each filled circle refers to one spaxel in each galaxy  and is colour-coded with the magnitude of velocity gradient (v$_{\rm grad}$, Equation \ref{eqn:grad}).
 The crosses and asterisks in different colours refer to the 11 actively star-forming galaxies at {\it z} $\sim$ 2 in \citet{Lehnert09}.       
 The solid black curves show  $\sigma \propto (\epsilon \dot{E})^{1/2}$, where $\dot{E}$ is the energy injection due to star formation, and  $\epsilon$ is the coupling efficiency of  the  energy injected into the ISM. The dashed curves show $\sigma \propto (\epsilon \dot{E})^{1/3}$ assuming that velocity dispersions correspond to  energy dissipation due to turbulent motions. The red solid curve shows the velocity dispersion of a 10$^8$ M$_\odot$ clump assuming a simple Jeans relation. A zoom-in of our SAMI galaxies with a logarithmic  y-axis is also shown here. All of the models have included the typical thermal broadening of H$\alpha$ of 12 km s$^{-1}$. The error bars in the zoom-in figure show the maximum range of the thermal broadening of 10 --15 km s$^{-1}$. The colourbar on the left shows the magnitude of the velocity gradient (v$\rm _g$). }
\label{fig:SFR_vdisp1}
\end{figure*}

%----------------------------------------------- end of figure 3-1 -------------------------------------------------------

%------------------------------------------- figure 3-2: genzel overplot -------------------------------------------------------
\begin{figure}
\centering
	\includegraphics[width=1.\columnwidth]{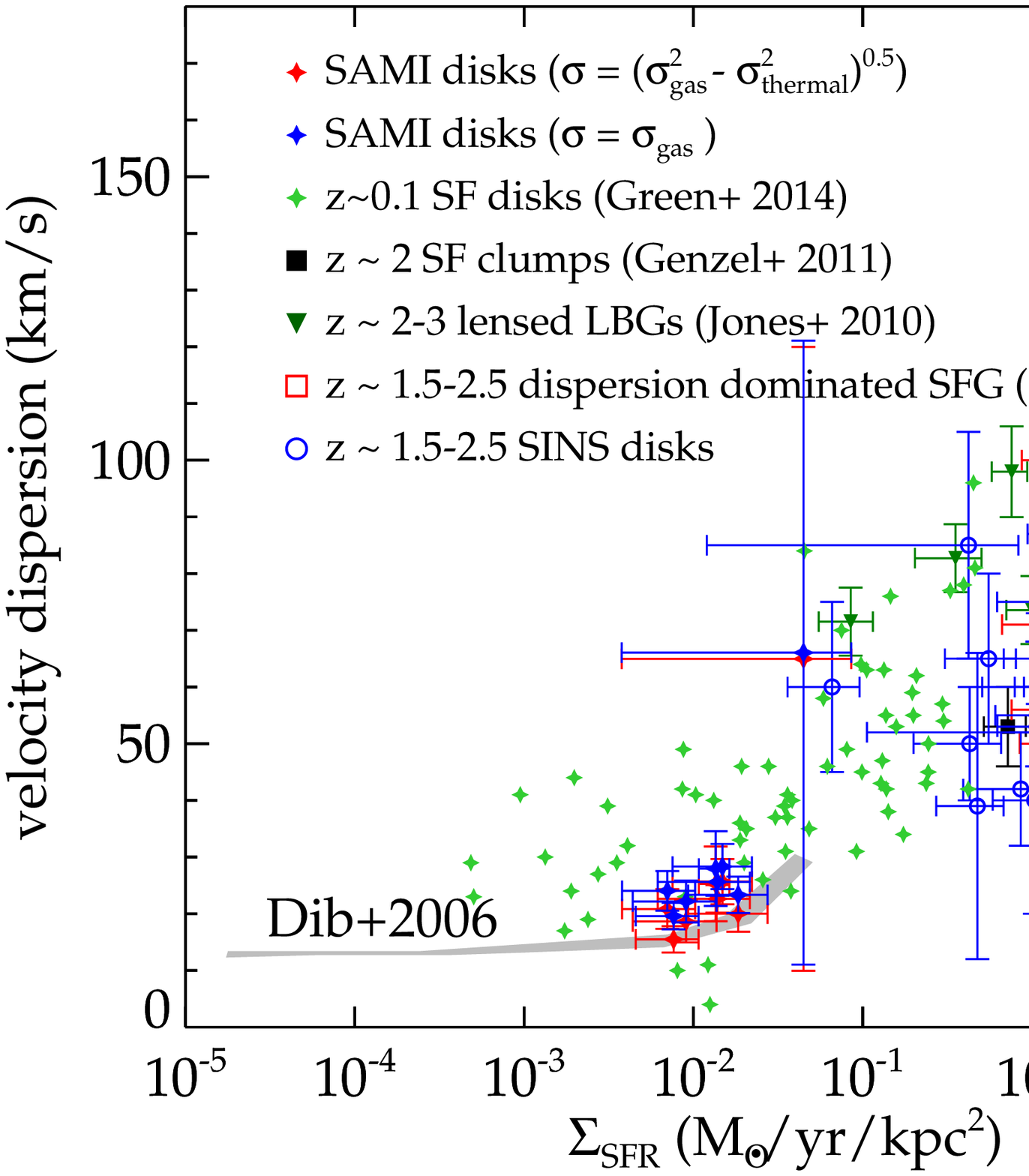}
    \caption{ Global dependence of $\sigma_{\rm gas}$ on $\Sigma_{\rm SFR}$.
     Our eight SAMI galaxies compared to   local high H$\alpha$ luminosity galaxies from \citet{Green14} and {\it z} > 1 star-forming galaxies and clumps (see Section \ref{sec:global}  for further details). Each blue (red) filled diamond shows one entire  galaxy in our sample including (excluding)  the contribution from thermal broadening \citep[$\sigma_{\rm thermal}$ $\sim$ 12 km s$^{-1}$, ][]{Glazebrook13}.  For the measurement of $\sigma_{\rm gas}$ and $\Sigma_{\rm SFR}$, see footnotes in Table \ref{tab:properties}. Green diamonds refer to the H$\alpha$ luminous  galaxies in \citet{Green14}.  The black filled squares, dark green triangles, red open squares and blue open circles refer to the {\it z} > 1 star-forming galaxies and clumps. The  grey contour denotes the distribution of local star-forming galaxies with gas velocity dispersion derived from HI \citep{Dib06}, and we include the intrinsic thermal broadening of 12 km s$^{-1}$ here. }
\label{fig:SFR_vdisp2}
\end{figure}

%----------------------------------------------- end of figure 3-2 -------------------------------------------------------

After removing the spaxels with high $ v_{\rm grad}$ (keeping those with $\sigma_{\rm gas}$ > 2 $ v_{\rm grad}$), we compare the spatially resolved relation between $\sigma_{\rm gas}$ and $\Sigma_{\rm SFR}$ in our local galaxies with that in high redshift galaxies \citep{Lehnert09} in Figure \ref{fig:SFR_vdisp1}. We also insert a zoom-in of our SAMI data with a logarithmic  y-axis.
%Here each point refers to one spaxel (i.e., one grid in the data cubes described in Section \ref{sec:data1}
%, denoting  $\rm 0.5\arcsec \times 0.5\arcsec$,  a physical scale of $\sim$ 0.5 $\times$ 0.5 kpc$^2$ 
%) in each galaxy and is colour-coded with the magnitude of  velocity shear (v$_{\rm shear}$, Equation \ref{eqn:grad}) at the position. Only spaxels with signal-to-noise ratio ( SNR = $\rm \frac{SFR}{SFR\_{err}}$ ) > 30 are displayed in Figure \ref{fig:SFR_vdisp}. 

As shown by the filled circles, the velocity dispersion of our local star-forming galaxies is almost constant around 20 km s$^{-1}$, covering a $\Sigma_{\rm SFR}$ range of over an order of magnitude. As  discussed in Section \ref{sec:snr}, SAMI is able to detect $\rm \sigma_{gas}$ below 15 km s$^{-1}$ for a sufficiently high S/N (here 34),  therefore the cut-off at $\sim$15 km s$^{-1}$ represents a physical cut-off. The flat and tight distribution of  $\sigma_{\rm gas}$  in our local star-forming galaxies does not present an obvious correlation with $\Sigma_{\rm SFR}$.
There are  some outliers with $\sigma_{\rm gas}$ > 40 km s$^{-1}$ that may result from a shocked component in 508421 as mentioned in  Section \ref{sec:sample} and \ref{sec:beamsmearing}. 

We also connect the  behaviour of the gas in star-forming galaxies from low-{\it z} to high-{\it z} by superimposing our data onto the results in Figure 12 from \citet{Lehnert09}. The crosses and asterisks in different colours refer to the 13 actively star-forming galaxies at {\it z} $\sim$ 2 in \citet{Lehnert09}. %The curves denote different models to be discussed in the next subsection (Section \ref{sec:model}).
%i.e., higher $\rm \sigma_{gas}$ at higher $\Sigma_{\rm SFR}$,
The high redshift galaxies in  \citet{Lehnert09}  show higher velocity dispersion with larger scatter than our local star-forming galaxies. They are also more or less positively correlated with  $\Sigma_{\rm SFR}$,  however, with substantial scatter. This may indicate that  star formation feedback plays a more important role  in high redshift star-forming galaxies than in their local counterparts. High redshift star-forming galaxies, as shown  in \citet{Genzel11, Genzel14}, are mostly irregular and clumpy  while  local galaxies are  stable and rotationally supported, as seen in Figure \ref{fig:images}. Therefore,  juvenile high redshift galaxies could be more easily affected by their intense star formation activity and  mature local galaxies could be more sensitive to galactic-scale dynamics like cloud-cloud collisions, galactic shear, self-gravity, and magnetorotational instability than to star formation feedback. We discuss this further in Section \ref{sec:discussion}.

\paragraph{Theoretical models for $\sigma_{\rm gas}$}

%{\bf Theoretical models for $\bf \sigma_{\bf gas}$}to the SN energy from disk star formation,
The curves in  Figure \ref{fig:SFR_vdisp1} denote different models proposed by \citet{Lehnert09}  to explain the relation between  $\sigma_{\rm gas}$ and $\Sigma_{\rm SFR}$.  

If the dissipated energy comes from star formation, and star formation induces high pressures, then a simple scaling relation is expected. This is indicated by the solid black curves, in the form  $\sigma \propto (\epsilon \dot{E})^{1/2}$, where $ \dot{E}$ is the energy injection rate due to star formation, and $\epsilon$ is the coupling efficiency between the injected energy and the ISM. According to \citet{Dib06}, when modelling the ISM with a coupling efficiency of 25\% (a conservative value),  quiescent galaxies may switch to a starburst mode at $\Sigma_{\rm SFR}$ = 10$^{-2.5}$ to 10$^{-2}$ M$_{\odot}$ yr$^{-1}$ kpc$^{-2}$. The bottom two black solid lines are derived from such models using these two values, showing $\rm \sigma_{\rm gas} = 100 \Sigma_{\rm SFR}^{1/2}$ and $\rm \sigma_{\rm gas} = 140 \Sigma_{\rm SFR}^{1/2}$ ($\sigma_{\rm gas}$ is in km s$^{-1}$,  $\Sigma_{\rm SFR}$ is in M$_\odot$ yr$^{-1}$ kpc$^{-2}$, the same below).  The third curve at the top  shows $\rm \sigma_{\rm gas} = 240 \Sigma_{\rm SFR}^{1/2}$, using coupling efficiencies of 100\%  (an extreme and unrealistic value).

If the energy is dissipated through incompressible turbulence, another scaling relation would be expected.
This is shown by the dashed curves in the form of $\sigma_{\rm gas} \propto (\epsilon \dot{E})^{1/3}$ where $\dot{E}$ is the  energy dissipated through turbulence. The two black dashed curves show $\rm \sigma_{\rm gas} = 80 \Sigma_{\rm SFR}^{1/3}$ and $\rm \sigma_{\rm gas} = 130 \Sigma_{\rm SFR}^{1/3}$, using coupling efficiencies of 25\%, 100\%, and a primary injection scale of 1 kpc.

If the turbulence is powered by gravity, assuming a simple Jeans relationship between mass and velocity dispersion, \citet{Lehnert09} derived the relation in the form of $\sigma_{\rm gas}$ $\sim$ $M_{J}^{1/4}G^{1/2}\Sigma_{\rm gas}^{1/4}$ = 54 km s$^{-1}$ $M_{J,9}^{1/4}\Sigma_{\rm SFR}^{0.18}$, where {\it G} is the gravitational constant, $\Sigma_{\rm gas}$ is the gas mass surface density in M$_{\odot}$ pc$^{-2}$, $M_{J,9}$ is the Jeans mass in units of 10$^{9}$ M$_{\odot}$ and $\Sigma_{\rm SFR}$ is in M$_{\odot}$ kpc$^{-2}$ yr$^{-1}$.
The red solid curve shows the velocity dispersion as a function of $\Sigma_{\rm SFR}$ of a 10$^8$ $\rm M_\odot$ giant molecular cloud (GMC). Given the mass of our local star-forming galaxies are similar to the Milky Way,  there is not any molecular cloud more massive than 10$^8$ $\rm M_\odot$ \citep{Roman-Duval10}, which means that the red curve represents an upper limit for the velocity dispersion obtained via the Jeans relation proposed by \citet{Lehnert09}

Due to a characteristic temperature of 10$^4$ K \citep{Andrews13}, the H$\alpha$ emission line has a typical thermal broadening of $\sim$ 12 km s$^{-1}$. In addition, the temperature distribution in an HII region is not uniform, so we estimate a maximum  range of 10 -- 15 km s$^{-1}$ \citep{Andrews13}.
Here in all of the models we include the intrinsic thermal broadening of 12 km s$^{-1}$, but we also display the maximum range with the error-bars in the zoom-in figure.

\bigskip

None of the models  can properly explain our local star-forming galaxies.  
$\sigma \propto (\epsilon \dot{E})^{1/2}$ and $\sigma \propto (\epsilon \dot{E})^{1/3}$ are consistent with some of the spaxels in our SAMI galaxies. However,   all of the relations predict lower velocity dispersion than seen in a significant number of spaxels.  Most of the data points of our galaxies are significantly above the bottom two black solid lines and the bottom black dashed line, which correspond to a realistic coupling efficiency of 25\% \citep{Dib06} in the two models.  Even if we consider the most extreme (and unrealistic) cases with coupling efficiencies of  100\%, as shown by the top black solid line and the top black dashed line, the datapoints at the lower $\Sigma_{\rm SFR}$ end cannot be well explained. The same is true with the Jeans instability relation. Even if we assume extreme GMC masses as high as as 10$^8$ $\rm M_\odot$, it still underestimates the velocity dispersion of our galaxies. Moreover,  the  error bars in the zoom-in figure  indicate that the uncertainty induced by the thermal broadening has minor influence on the distribution of the models.
Therefore, simply considering energy injection from star formation, dissipation of incompressible turbulence, or release of gravitational energy alone as the source of velocity dispersion is not enough to explain the distribution of gas velocity dispersion in local star-forming galaxies.

\subsubsection{Global (galaxy-averaged) analysis}
\label{sec:global}
In Figure \ref{fig:SFR_vdisp2}, we look at the global behaviour  of the gas in star-forming galaxies. We include the local H$\alpha$ luminous galaxies  from \citet{Green14}  as   green diamonds, and the {\it z} > 1 star-forming galaxies and star-forming clumps with good data quality\footnote{which includes --  a) {\it z} $\sim$ 2 star-forming clumps \citep{Genzel11} as black filled squares;  b) {\it z} $\sim$ 2--3 low mass ($\sim$ 10$^9$ M$_{\odot}$) lensed star-forming galaxies \citep{Jones10} as dark green triangles;  c) {\it z} $\sim$ 1.5--3 low mass ($\sim$ 0.3--3 $\times$10$^{10}$ M$_{\odot}$), compact but well-resolved with adaptive optics "dispersion dominated" star-forming galaxies \citep{Law09} as red open squares; {\bf d)} {\it z} $\sim$ 1.5--2.5 disks from the SINS survey \citep{Cresci09, Forster09} as blue open circles.}, for comparison. The area used to calculate the $\Sigma_{\rm SFR}$ of galaxies in \citet{Green14}  are  from the radii based on an exponential model fit. Given their various sizes, they span a very large range of $\Sigma_{\rm SFR}$. The global $\Sigma_{\rm SFR}$s and velocity dispersions of our eight galaxies are listed in Table \ref{tab:properties}. The global $\Sigma_{\rm SFR}$s and velocity dispersions are the flux weighted averages of spaxels within individual galaxies. Only the spaxels with $\sigma_{\rm gas}$ > 2 $v_{\rm grad}$ are considered in the measurement. All these works  derive velocity dispersion from the H$\alpha$ emission lines. 
The  grey contour  denotes the distribution of local star forming galaxies with $\rm \sigma_{gas}$ derived from HI \citep{Dib06}, and we include the intrinsic thermal broadening of 12 km s$^{-1}$ here.
The difference in velocity dispersions between HI and H$\alpha$ comes from the thermal broadening of warm ionised gas at $\sim$ 10$^{4}$ K. 

 %---------------------------------------- table 3: correlation coeffecient ------------------------------------------
\begin{table}
\centering
%\scriptsize
\caption{\label{tab:Spearman} Spearman correlation coefficient ($r_s$).}
\begin{threeparttable}
\begin{tabular}{lcccccc}
\hline
&\multicolumn{2}{c}{$\rm \sigma_{gas}$ vs. $\rm \Sigma_{SFR}$} \\
\cline{2-3}
&$r_s$&P\tnote{$\ast$}\\
\hline
SAMI + Green\tnote{$\dagger$} 								 	&0.72  	&$\ll$0.01 \\
SAMI + high-{\it z}\tnote{$\ddagger$} 							&0.53 	&$\ll$0.01   \\
SAMI + Green\tnote{$\dagger$} + high-{\it z}\tnote{$\ddagger$}	&0.76 	&$\ll$0.01 \\
\hline
\end{tabular}
\begin{tablenotes}
        \item[$\dagger$]H$\alpha$ luminous galaxies in \citet{Green14}.
        \item[$\ddagger$]{\it z} >1 star-forming galaxies and star-forming clumps. See Section \ref{sec:global} for further detail.
        \item[$\ast$]Significance level of the Spearman correlation coefficient.
      \end{tablenotes}
      \end{threeparttable}
\end{table}
%---------------------------------------- end of table 3 ------------------------------------------

Comparing our eight SAMI galaxies to the high redshift galaxies, we see similar behaviour to the spatially resolved data (c.f. Figure \ref{fig:SFR_vdisp1}). When taking the 67 H$\alpha$ luminous galaxies from \citet{Green14} into account, we find that it can well connect the local and high-redshift galaxies. In  Figure \ref{fig:SFR_vdisp2}, we see that their distribution at the lower $\Sigma_{\rm SFR}$ end agrees with  the distribution of our eight galaxies and the higher $\Sigma_{\rm SFR}$ end  follows the distribution of the high  redshift galaxies as well. \citet{Green14}'s local star-forming galaxies are chosen to have  similar properties to the high redshift galaxies, such as  high gas fraction and  high H$\alpha$ luminosities ( $\gtrsim 10^{42}$ erg s$^{-1}$).  
Note that we calculate the $\Sigma_{\rm SFR}$ of galaxies in \citet{Green14} without flux weighting, so the $\Sigma_{\rm SFR}$ may be underestimated compared to other galaxies.

\bigskip
We measure the Spearman correlation coefficients between gas velocity dispersion and $\Sigma_{\rm SFR}$, for three different groups of the three sets of data. The results are listed in Table \ref{tab:Spearman}. Our eight SAMI galaxies together with \citet{Green14}'s H$\alpha$ luminous galaxies show strong correlation between velocity dispersion and $\Sigma_{\rm SFR}$ ($r_{s}$ = 0.72). When we include the high-{\it z} galaxies, the correlations become a bit stronger ($r_{s}$ = 0.76). However, when  combining our eight SAMI galaxies with only the high-{\it z} galaxies, the correlation becomes moderate ($r_{s}$ = 0.53).  Note that the global $\Sigma_{\rm SFR}$ are flux weighted. Therefore this result comes from relatively more active star-forming regions, which is not contrary to the conclusion  we draw from the spatially resolved analysis in Section \ref{sec:sfrsigma}. Stellar feedback alone is insufficient to drive the observed $\sigma_{\rm gas}$, especially at low SFR surface density (Section \ref{sec:sfrsigma}), while it can become dominant  when SFR surface density is high enough (Section \ref{sec:global}).
%{\bf Figure \ref{fig:SFR_vdisp1} and  Figure \ref{fig:SFR_vdisp2} gives opposite conclusions suggesting that different energy sources dominate on sub-kpc scales and across individual galaxies separately. }

\section{discussion}
\label{sec:discussion}

\subsection{Main driver(s) of velocity dispersion}
\label{sec:drivers}
As mentioned in Section \ref{sec:sfrsigma},   the flat and elevated (compared to the model predictions) distribution of spaxels of our SAMI galaxies shown in Figure \ref{fig:SFR_vdisp1} indicates that the star formation feedback is unlikely to dominate  the gas velocity dispersion at the scale of $\sim$ 0.5 $\times$ 0.5 kpc$^2$. Thus, additional drivers of turbulence must be acting in these local, low $\Sigma_{\rm SFR}$ galaxies.  As indicated by the flatness of the distribution, such sources need to be common among galaxies and not vary much within galaxies. 

The drivers of turbulence can either compress the gas (compression processes) or directly excite solenoidal motions of gas (stirring processes) \citep{Federrath16b, Federrath17}. 

Stellar feedback like stellar winds of OB stars and Wolf Rayet stars, supernova explosions, as well as accretion processes (such as accretion onto a galaxy) and gravitational contraction  are able to compression the gas, and then increase gas velocity dispersion and induce star formation at the same time. Therefore, SFR (or $\Sigma_{\rm SFR}$) may not be directly related to gas velocity dispersion. Both SFR (or $\Sigma_{\rm SFR}$) and gas velocity dispersion can be affected simultaneously by the same sources (e.g. accretion, gravity etc.), but to different degree.

\citet{Krumholz16} proposed that the turbulence in the ISM is driven by gravity rather than  stellar feedback. The higher (than expected) velocity dispersion in Figure \ref{fig:SFR_vdisp1} may be due to the release of gravitational energy.  However,  \citet{Krumholz16}'s conclusion comes from the constraint of rapid star-forming, high velocity dispersion galaxies. Their models do not show any apparent difference at low  velocity dispersion. Moreover, they investigate the relation with SFR rather than $\Sigma_{\rm SFR}$ and on entire galaxies rather than on spatially resolved regions. 
On the other hand, \citet{Lehnert09}'s models in Figure \ref{fig:SFR_vdisp1} reveal that -- (i) self-gravity  is not sufficient even if  the model (red solid curve) adopts an extremely massive the star-forming clump as 10$^8$ M$_{\odot}$, which is too big for local star-forming galaxies; (ii) turbulence can be driven by the bulk motions induced by energy injection from star formation, and then cascade, redistribute and dissipate the energy down to smallest scales.  Star formation powers the turbulence of galaxies with high velocity dispersions in complex ways, but star formation alone is insufficient to explain the gas velocity dispersions in our local star-forming galaxies.  Generally, both gravity and star formation can power the turbulence but  dominate at different  redshifts and/or  in different environments.

Stirring processes like galactic shear, MRI, and jets/outflows can induce solenoidal motions, i.e., increase the velocity dispersion, but suppress the star formation.  
Shear is a typical driver of the turbulence in the centres of our galaxy and possibly other galaxies \citep{KK15, Kruijssen16, Federrath16b, Federrath17}. MRI requires a combination of rotation and magnetic fields. Compared to compressive stirring mechanisms, solenoidal stirring processes have less influence on the density distribution \citep{Federrath08, Federrath10}. Therefore, solenoidal driving mechanisms reduce the SFR compared to compressive sources of $\sigma_{\rm gas}$ \citep{Federrath12}. 
Solenoidal drivers (such as MRI and shear) may be able to provide an explanation for the distribution of our SAMI galaxies in Figure \ref{fig:SFR_vdisp1}: suppressed $\Sigma_{\rm SFR}$ but relatively higher gas velocity dispersion than expected from star formation feedback.

Note that the velocity modes (solenoidal and compressible) do not grow independently of one another (for a detailed analysis in the case of compressive and solenoidal driving of the turbulence, see \citet[][Fig. 14]{Federrath10} and \citet[][Fig. 3 bottom panel]{Federrath11}). However, what we are referring to here when we talk about solenoidal and compressive modes, are the modes in the acceleration field (not the resulting velocity field) that drives the turbulence. A summary of the differences and implications of solenoidally- and compressively-driven turbulence and their implications for star formation is presented in \citet{Federrath17} and the main theoretical framework as well as comparison to observations are presented in \citet{Federrath12}.

\subsection{Caveats}
\label{sec:limit}

\subsubsection{The medium being observed} 
H$\alpha$ emission traces ionized gas and gives information on the turbulences in HII regions. We may miss out the tracers showing the star formation feedback and effects of turbulence driven by galaxy dynamics (such as shear) by just analysing the ionized gas (and not including the atomic and molecular phases).  Studies such as \citet{Stilp13} find two component fits are necessary for HI lines in nearby galaxies, and the broad components are mainly related to the star-formation intensity, which support the idea of star formation supporting   the turbulence and gravitational instability/shear.
Here we are limited to strong emission line data, but we have applied for ALMA time to follow-up a subset of our galaxies to study the velocity dispersion in the cold/molecular gas as well.%To do a complete study of the energy sources of turbulent velocity dispersions,   direct detection of cold gas like HI, H$_2$, or CO etc. is also needed. So far, the spectral resolution of HI detection is not good enough to get accurate velocity dispersions. Our team has proposed to ALMA to obtain CO velocity dispersion maps of a sample of star-forming galaxies extracted from SAMI for a further study of the regulations on the turbulence of ISM.

\subsubsection{Removing the inner regions of the galaxies}
\label{sec:inner}

In this paper, circum-nuclear regions of our galaxies are  removed because of the strong influence of beam-smearing. However, circum-nuclear regions are very interesting and important in order to develop a complete picture of turbulence injection across a galaxy. The influence of the rotation curve could be especially important in circum-nuclear regions, providing a strong test of the gravitational shear/instability arguments. The broadest molecular lines are also found
in the circum-nuclear gas of galaxies suggesting that star formation is playing a crucial role there \citep[e.g.][]{Wilson11}. Here we chose to exclude the inner regions, where beam smearing significantly affects our data. Future studies that attempt to correct for beam smearing will be needed to investigate the turbulence in the circum-nuclear regions in detail.

\subsubsection{Removing the galaxies that show any evidence for shocks}
Shock driven turbulence contains important information about the connection between star formation feedback and turbulence.
However, it is difficult to disentangle the contribution of star formation from possible AGN activity to extract actual SFRs accurately from the shock-heated gas. We would  likely overestimate SFRs and thus overestimate the effect of star formation feedback. Thus, in a conservative way, we chose not to include any galaxies with signatures of shock-excited emission and focus on purely star-forming galaxies.

\subsubsection{Dependence of velocity dispersion on rotational velocity}

\begin{figure}
\centering 
	\includegraphics[width=1\columnwidth]{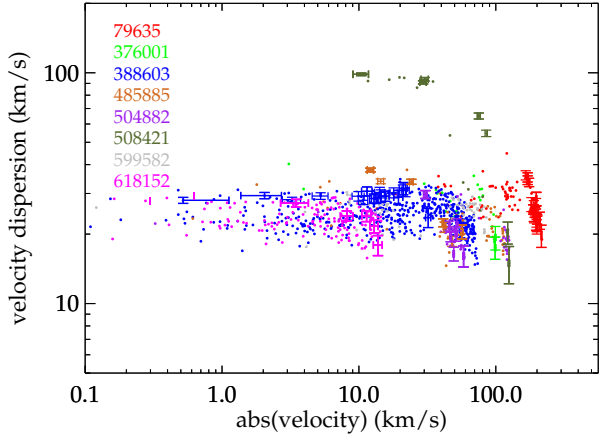}
	\caption{Velocity dispersion as a function of velocity. Spaxels along the major axis are plotted with errorbars. Points in each individual galaxy are labeled by colour.}
\label{fig:v_vdisp}
\end{figure}

Figure \ref{fig:v_vdisp} shows the dependence of velocity dispersion on rotational velocity. Overall, we see no strong dependence. However, we  see some trends of decreasing velocity dispersion  with increasing velocity, especially along the major axes (labeled with errorbars).  The beam smearing effect would be higher in the center than the outside. For our galaxies, we have already removed the highly affected spaxels, thus the circum-nuclear regions are removed (Section \ref{sec:inner}). %Note that we are not correcting beam smearing, there might remain slight dependence of velocity dispersion on rotational velocity. But given we are using a strict beam smearing criterion, we assume such trend are more likely to be physical -- increasing velocity dispersion (closer to the center) is because of increasing SFR or of enhanced shear.

\section{conclusion}
\label{sec:conclusion}
Random, turbulent motions are important in regulating the star formation in the galaxies, but the energy source of  turbulent motions remains unclear. We investigate the random motions of the ionised gas in local star-forming galaxies.
After very strict selection to avoid the possible contamination by AGN, shocks and outflows, we find eight SAMI galaxies satisfying the pure star-forming criteria based on emission line diagnostic diagrams.   We  minimise  the influence of beam smearing by removing the spaxels with $\sigma_{\rm gas}$ < 2 $v_{\rm grad}$ before further analysis (Figure \ref{fig:grad_contrast}). The spatially resolved images and  high spectral resolution of the SAMI Galaxy Survey shows that, on scales of 0.5 $\times$ 0.5 kpc$^2$, turbulence within local star-forming galaxies is not exclusively driven by star formation feedback. The flat but elevated (compared to model predictions) distribution of gas velocity dispersion as a function of $\Sigma_{\rm SFR}$ (Figure \ref{fig:SFR_vdisp1}) implies that there must be some additional energy source(s) besides star formation feedback,  especially at the low $\Sigma_{\rm SFR}$ end. Such source(s)  need to be common among local star-forming galaxies and do not vary spatially across galaxies. The difference with the high-redshift galaxies and H$\alpha$ luminous galaxies (Figures \ref{fig:SFR_vdisp1} and \ref{fig:SFR_vdisp2}) indicates that the low SFR in these local galaxies  is too weak to explain the random motions of the ionised gas. Juvenile high redshift galaxies could be more sensitive to their intense star formation activity while mature local galaxies could be more influenced by galactic-scale dynamics like gravity, galactic shear,  and MRI  than by star formation feedback.

\section*{Acknowledgements}

%%%Wenjia NJU
L.Z.  and Y.S.  acknowledge support for this work from the National Natural Science Foundation of China (NSFC grant 11373021) and the Excellent Youth Foundation of the Jiangsu Scientific Committee (grant BK20150014). 
%%%%%%Christoph
C.F. gratefully acknowledges funding provided by the Australian Research Council's Discovery Projects (grants DP150104329 and DP170100603).
%%%%%%%% ANNE
Support for AMM is provided by NASA through Hubble Fellowship grant \#HST-HF2-51377 awarded by the Space Telescope Science Institute, which is operated by the Association of Universities for Research in Astronomy, Inc., for NASA, under contract NAS5-26555. 
We thank R.Genzel, N.F{\"o}rster Schreiber, M.D.Lehnert, and N.P.H. Nesvadba for their support on providing high-{\it z} measurements displayed in  this work. 
%%%%%
M.S.O. acknowledges the funding support from the Australian Research Council through a Future Fellowship (FT140100255).
%%%
B.C. acknowledges support from the Australian Research Council's Future Fellowship (FT120100660) funding scheme.
%%%%%
S.B. acknowledges the funding support from the Australian Research Council through a Future Fellowship (FT140101166).
%%%%%%
SFS thanks PAPIIT-DGAPA-IA101217 project and CONACYT-IA-180125.
%%%%%%%
The authors would like to thank the anonymous reviewer for his/her helpful and constructive comments that greatly contributed to improving the final version of the paper.

%%%%%% SAMI FROM WEB
The SAMI Galaxy Survey is based on observations made at the Anglo-Australian Telescope. The Sydney-AAO Multi-object Integral field spectrograph (SAMI) was developed jointly by the University of Sydney and the Australian Astronomical Observatory. The SAMI input catalogue is based on data taken from the Sloan Digital Sky Survey, the GAMA Survey and the VST ATLAS Survey. The SAMI Galaxy Survey is funded by the Australian Research Council Centre of Excellence for All-sky Astrophysics (CAASTRO), through project number CE110001020, and other participating institutions. The SAMI Galaxy Survey website is \href{url}{http://sami-survey.org/}.

%%%%%%%%%%%%%%%%%%%%%%%%%%%%%%%%%%%%%%%%%%%%%%%%%%



%\bibliographystyle{mnras}


\bibliographystyle{mnras}
\bibliography{bibtem}

\begin{thebibliography}{}
\makeatletter
\relax
\def\mn@urlcharsother{\let\do\@makeother \do\$\do\&\do\#\do\^\do\_\do\%\do\~}
\def\mn@doi{\begingroup\mn@urlcharsother \@ifnextchar [ {\mn@doi@}
  {\mn@doi@[]}}
\def\mn@doi@[#1]#2{\def\@tempa{#1}\ifx\@tempa\@empty \href
  {http://dx.doi.org/#2} {doi:#2}\else \href {http://dx.doi.org/#2} {#1}\fi
  \endgroup}
\def\mn@eprint#1#2{\mn@eprint@#1:#2::\@nil}
\def\mn@eprint@arXiv#1{\href {http://arxiv.org/abs/#1} {{\tt arXiv:#1}}}
\def\mn@eprint@dblp#1{\href {http://dblp.uni-trier.de/rec/bibtex/#1.xml}
  {dblp:#1}}
\def\mn@eprint@#1:#2:#3:#4\@nil{\def\@tempa {#1}\def\@tempb {#2}\def\@tempc
  {#3}\ifx \@tempc \@empty \let \@tempc \@tempb \let \@tempb \@tempa \fi \ifx
  \@tempb \@empty \def\@tempb {arXiv}\fi \@ifundefined
  {mn@eprint@\@tempb}{\@tempb:\@tempc}{\expandafter \expandafter \csname
  mn@eprint@\@tempb\endcsname \expandafter{\@tempc}}}

\bibitem[\protect\citeauthoryear{{Allen} et~al.,}{{Allen}
  et~al.}{2015}]{Allen15}
{Allen} J.~T.,  et~al., 2015, \mn@doi [\mnras] {10.1093/mnras/stu2057}, \href
  {http://adsabs.harvard.edu/abs/2015MNRAS.446.1567A} {446, 1567}

\bibitem[\protect\citeauthoryear{{Andrews} \& {Martini}}{{Andrews} \&
  {Martini}}{2013}]{Andrews13}
{Andrews} B.~H.,  {Martini} P.,  2013, \mn@doi [\apj]
  {10.1088/0004-637X/765/2/140}, \href
  {http://adsabs.harvard.edu/abs/2013ApJ...765..140A} {765, 140}

\bibitem[\protect\citeauthoryear{{Baldry} et~al.,}{{Baldry}
  et~al.}{2012}]{Baldry12}
{Baldry} I.~K.,  et~al., 2012, \mn@doi [\mnras]
  {10.1111/j.1365-2966.2012.20340.x}, \href
  {http://adsabs.harvard.edu/abs/2012MNRAS.421..621B} {421, 621}

\bibitem[\protect\citeauthoryear{{Baldwin}, {Phillips}  \&
  {Terlevich}}{{Baldwin} et~al.}{1981}]{Baldwin81}
{Baldwin} J.~A.,  {Phillips} M.~M.,   {Terlevich} R.,  1981, \mn@doi [\pasp]
  {10.1086/130766}, \href {http://adsabs.harvard.edu/abs/1981PASP...93....5B}
  {93, 5}

\bibitem[\protect\citeauthoryear{{Bassett} et~al.,}{{Bassett}
  et~al.}{2014}]{Bassett14}
{Bassett} R.,  et~al., 2014, \mn@doi [\mnras] {10.1093/mnras/stu1029}, \href
  {http://adsabs.harvard.edu/abs/2014MNRAS.442.3206B} {442, 3206}

\bibitem[\protect\citeauthoryear{{Baugh}, {Cole}  \& {Frenk}}{{Baugh}
  et~al.}{1996}]{Baugh96}
{Baugh} C.~M.,  {Cole} S.,   {Frenk} C.~S.,  1996, \mn@doi [\mnras]
  {10.1093/mnras/283.4.1361}, \href
  {http://adsabs.harvard.edu/abs/1996MNRAS.283.1361B} {283, 1361}

\bibitem[\protect\citeauthoryear{{Bland-Hawthorn} et~al.,}{{Bland-Hawthorn}
  et~al.}{2011}]{BlandHawthorn11}
{Bland-Hawthorn} J.,  et~al., 2011, \mn@doi [Optics Express]
  {10.1364/OE.19.002649}, \href
  {http://adsabs.harvard.edu/abs/2011OExpr..19.2649B} {19, 2649}

\bibitem[\protect\citeauthoryear{{Bryant}, {Bland-Hawthorn}, {Fogarty},
  {Lawrence}  \& {Croom}}{{Bryant} et~al.}{2014}]{Bryant14}
{Bryant} J.~J.,  {Bland-Hawthorn} J.,  {Fogarty} L.~M.~R.,  {Lawrence} J.~S.,
  {Croom} S.~M.,  2014, \mn@doi [\mnras] {10.1093/mnras/stt2254}, \href
  {http://adsabs.harvard.edu/abs/2014MNRAS.438..869B} {438, 869}

\bibitem[\protect\citeauthoryear{{Bryant} et~al.,}{{Bryant}
  et~al.}{2015}]{Bryant15}
{Bryant} J.~J.,  et~al., 2015, \mn@doi [\mnras] {10.1093/mnras/stu2635}, \href
  {http://adsabs.harvard.edu/abs/2015MNRAS.447.2857B} {447, 2857}

\bibitem[\protect\citeauthoryear{{Bundy} et~al.,}{{Bundy}
  et~al.}{2015}]{Bundy15}
{Bundy} K.,  et~al., 2015, \mn@doi [\apj] {10.1088/0004-637X/798/1/7}, \href
  {http://adsabs.harvard.edu/abs/2015ApJ...798....7B} {798, 7}

\bibitem[\protect\citeauthoryear{{Cresci} et~al.,}{{Cresci}
  et~al.}{2009}]{Cresci09}
{Cresci} G.,  et~al., 2009, \mn@doi [\apj] {10.1088/0004-637X/697/1/115}, \href
  {http://adsabs.harvard.edu/abs/2009ApJ...697..115C} {697, 115}

\bibitem[\protect\citeauthoryear{{Croom} et~al.,}{{Croom}
  et~al.}{2012}]{Croom12}
{Croom} S.~M.,  et~al., 2012, \mn@doi [\mnras]
  {10.1111/j.1365-2966.2011.20365.x}, \href
  {http://adsabs.harvard.edu/abs/2012MNRAS.421..872C} {421, 872}

\bibitem[\protect\citeauthoryear{{Daddi} et~al.,}{{Daddi}
  et~al.}{2010}]{Daddi10}
{Daddi} E.,  et~al., 2010, \mn@doi [\apj] {10.1088/0004-637X/713/1/686}, \href
  {http://adsabs.harvard.edu/abs/2010ApJ...713..686D} {713, 686}

\bibitem[\protect\citeauthoryear{{Dib}, {Bell}  \& {Burkert}}{{Dib}
  et~al.}{2006}]{Dib06}
{Dib} S.,  {Bell} E.,   {Burkert} A.,  2006, \mn@doi [\apj] {10.1086/498857},
  \href {http://adsabs.harvard.edu/abs/2006ApJ...638..797D} {638, 797}

\bibitem[\protect\citeauthoryear{{Elmegreen}}{{Elmegreen}}{2009}]{Elmegreen09}
{Elmegreen} B.~G.,  2009, in {Andersen} J.,  {Nordstr{\"o}ara} {m} B.,
  {Bland-Hawthorn} J.,  eds,  IAU Symposium Vol. 254, The Galaxy Disk in
  Cosmological Context. pp 289--300 (\mn@eprint {arXiv} {0810.5406}),
  \mn@doi{10.1017/S1743921308027713}

\bibitem[\protect\citeauthoryear{{Elmegreen} \& {Scalo}}{{Elmegreen} \&
  {Scalo}}{2004}]{Elmegreen04}
{Elmegreen} B.~G.,  {Scalo} J.,  2004, \mn@doi [\araa]
  {10.1146/annurev.astro.41.011802.094859}, \href
  {http://adsabs.harvard.edu/abs/2004ARA%26A..42..211E} {42, 211}

\bibitem[\protect\citeauthoryear{{Federrath} \& {Klessen}}{{Federrath} \&
  {Klessen}}{2012}]{Federrath12}
{Federrath} C.,  {Klessen} R.~S.,  2012, \mn@doi [\apj]
  {10.1088/0004-637X/761/2/156}, \href
  {http://adsabs.harvard.edu/abs/2012ApJ...761..156F} {761, 156}

\bibitem[\protect\citeauthoryear{{Federrath}, {Klessen}  \&
  {Schmidt}}{{Federrath} et~al.}{2008}]{Federrath08}
{Federrath} C.,  {Klessen} R.~S.,   {Schmidt} W.,  2008, \mn@doi [\apjl]
  {10.1086/595280}, \href {http://adsabs.harvard.edu/abs/2008ApJ...688L..79F}
  {688, L79}

\bibitem[\protect\citeauthoryear{{Federrath}, {Roman-Duval}, {Klessen},
  {Schmidt}  \& {Mac Low}}{{Federrath} et~al.}{2010}]{Federrath10}
{Federrath} C.,  {Roman-Duval} J.,  {Klessen} R.~S.,  {Schmidt} W.,   {Mac Low}
  M.-M.,  2010, \mn@doi [\aap] {10.1051/0004-6361/200912437}, \href
  {http://adsabs.harvard.edu/abs/2010A%26A...512A..81F} {512, A81}

\bibitem[\protect\citeauthoryear{Federrath, Chabrier, Schober, Banerjee,
  Klessen  \& Schleicher}{Federrath et~al.}{2011}]{Federrath11}
Federrath C.,  Chabrier G.,  Schober J.,  Banerjee R.,  Klessen R.~S.,
  Schleicher D. R.~G.,  2011, \mn@doi [Phys. Rev. Lett.]
  {10.1103/PhysRevLett.107.114504}, 107, 114504

\bibitem[\protect\citeauthoryear{{Federrath} et~al.,}{{Federrath}
  et~al.}{2016}]{Federrath16b}
{Federrath} C.,  et~al., 2016, \mn@doi [\apj] {10.3847/0004-637X/832/2/143},
  \href {http://adsabs.harvard.edu/abs/2016ApJ...832..143F} {832, 143}

\bibitem[\protect\citeauthoryear{{Federrath} et~al.,}{{Federrath}
  et~al.}{2017}]{Federrath17}
{Federrath} C.,  et~al., 2017, in {Crocker} R.~M.,  {Longmore} S.~N.,
  {Bicknell} G.~V.,  eds,  IAU Symposium Vol. 322, The Multi-Messenger
  Astrophysics of the Galactic Centre. pp 123--128 (\mn@eprint {arXiv}
  {1609.08726}), \mn@doi{10.1017/S1743921316012357}

\bibitem[\protect\citeauthoryear{{F{\"o}rster Schreiber} et~al.,}{{F{\"o}rster
  Schreiber} et~al.}{2009}]{Forster09}
{F{\"o}rster Schreiber} N.~M.,  et~al., 2009, \mn@doi [\apj]
  {10.1088/0004-637X/706/2/1364}, \href
  {http://adsabs.harvard.edu/abs/2009ApJ...706.1364F} {706, 1364}

\bibitem[\protect\citeauthoryear{{Genzel} et~al.,}{{Genzel}
  et~al.}{2008}]{Genzel08}
{Genzel} R.,  et~al., 2008, \mn@doi [\apj] {10.1086/591840}, \href
  {http://adsabs.harvard.edu/abs/2008ApJ...687...59G} {687, 59}

\bibitem[\protect\citeauthoryear{{Genzel} et~al.,}{{Genzel}
  et~al.}{2011}]{Genzel11}
{Genzel} R.,  et~al., 2011, \mn@doi [\apj] {10.1088/0004-637X/733/2/101}, \href
  {http://adsabs.harvard.edu/abs/2011ApJ...733..101G} {733, 101}

\bibitem[\protect\citeauthoryear{{Genzel} et~al.,}{{Genzel}
  et~al.}{2014}]{Genzel14}
{Genzel} R.,  et~al., 2014, \mn@doi [\apj] {10.1088/0004-637X/785/1/75}, \href
  {http://adsabs.harvard.edu/abs/2014ApJ...785...75G} {785, 75}

\bibitem[\protect\citeauthoryear{{Glazebrook}}{{Glazebrook}}{2013}]{Glazebrook13}
{Glazebrook} K.,  2013, \mn@doi [\pasa] {10.1017/pasa.2013.34}, \href
  {http://adsabs.harvard.edu/abs/2013PASA...30...56G} {30, e056}

\bibitem[\protect\citeauthoryear{{Green} et~al.,}{{Green}
  et~al.}{2010}]{Green10}
{Green} A.~W.,  et~al., 2010, \mn@doi [\nat] {10.1038/nature09452}, \href
  {http://adsabs.harvard.edu/abs/2010Natur.467..684G} {467, 684}

\bibitem[\protect\citeauthoryear{{Green} et~al.,}{{Green}
  et~al.}{2014}]{Green14}
{Green} A.~W.,  et~al., 2014, \mn@doi [\mnras] {10.1093/mnras/stt1882}, \href
  {http://adsabs.harvard.edu/abs/2014MNRAS.437.1070G} {437, 1070}

\bibitem[\protect\citeauthoryear{{Gritschneder}, {Naab}, {Walch}, {Burkert}  \&
  {Heitsch}}{{Gritschneder} et~al.}{2009}]{Gritschneder09}
{Gritschneder} M.,  {Naab} T.,  {Walch} S.,  {Burkert} A.,   {Heitsch} F.,
  2009, \mn@doi [\apjl] {10.1088/0004-637X/694/1/L26}, \href
  {http://adsabs.harvard.edu/abs/2009ApJ...694L..26G} {694, L26}

\bibitem[\protect\citeauthoryear{{Ho} et~al.,}{{Ho} et~al.}{2014}]{Ho14}
{Ho} I.-T.,  et~al., 2014, \mn@doi [\mnras] {10.1093/mnras/stu1653}, \href
  {http://adsabs.harvard.edu/abs/2014MNRAS.444.3894H} {444, 3894}

\bibitem[\protect\citeauthoryear{{Ho} et~al.,}{{Ho} et~al.}{2016}]{Ho16}
{Ho} I.,  et~al., 2016, preprint, \href
  {http://adsabs.harvard.edu/abs/2016arXiv160706561H} {} (\mn@eprint {arXiv}
  {1607.06561})

\bibitem[\protect\citeauthoryear{{Jones}, {Swinbank}, {Ellis}, {Richard}  \&
  {Stark}}{{Jones} et~al.}{2010}]{Jones10}
{Jones} T.~A.,  {Swinbank} A.~M.,  {Ellis} R.~S.,  {Richard} J.,   {Stark}
  D.~P.,  2010, \mn@doi [\mnras] {10.1111/j.1365-2966.2010.16378.x}, \href
  {http://adsabs.harvard.edu/abs/2010MNRAS.404.1247J} {404, 1247}

\bibitem[\protect\citeauthoryear{{Kelvin} et~al.,}{{Kelvin}
  et~al.}{2012}]{Kelvin12}
{Kelvin} L.~S.,  et~al., 2012, \mn@doi [\mnras]
  {10.1111/j.1365-2966.2012.20355.x}, \href
  {http://adsabs.harvard.edu/abs/2012MNRAS.421.1007K} {421, 1007}

\bibitem[\protect\citeauthoryear{{Kennicutt}, {Tamblyn}  \&
  {Congdon}}{{Kennicutt} et~al.}{1994}]{Kennicutt94a}
{Kennicutt} Jr. R.~C.,  {Tamblyn} P.,   {Congdon} C.~E.,  1994, \mn@doi [\apj]
  {10.1086/174790}, \href {http://adsabs.harvard.edu/abs/1994ApJ...435...22K}
  {435, 22}

\bibitem[\protect\citeauthoryear{{Kewley}, {Dopita}, {Sutherland}, {Heisler}
  \& {Trevena}}{{Kewley} et~al.}{2001}]{Kewley01}
{Kewley} L.~J.,  {Dopita} M.~A.,  {Sutherland} R.~S.,  {Heisler} C.~A.,
  {Trevena} J.,  2001, \mn@doi [\apj] {10.1086/321545}, \href
  {http://adsabs.harvard.edu/abs/2001ApJ...556..121K} {556, 121}

\bibitem[\protect\citeauthoryear{{Klessen} \& {Hennebelle}}{{Klessen} \&
  {Hennebelle}}{2010}]{Klessen1011}
{Klessen} R.~S.,  {Hennebelle} P.,  2010, \mn@doi [\aap]
  {10.1051/0004-6361/200913780}, \href
  {http://adsabs.harvard.edu/abs/2010A%26A...520A..17K} {520, A17}

\bibitem[\protect\citeauthoryear{{Kruijssen}}{{Kruijssen}}{2016}]{Kruijssen16}
{Kruijssen} J.~M.~D.,  2016, preprint, \href
  {http://adsabs.harvard.edu/abs/2016arXiv160908158K} {} (\mn@eprint {arXiv}
  {1609.08158})

\bibitem[\protect\citeauthoryear{{Krumholz} \& {Burkhart}}{{Krumholz} \&
  {Burkhart}}{2015}]{Krumholz15}
{Krumholz} M.~R.,  {Burkhart} B.,  2015, preprint, \href
  {http://adsabs.harvard.edu/abs/2015arXiv151203439K} {} (\mn@eprint {arXiv}
  {1512.03439})

\bibitem[\protect\citeauthoryear{{Krumholz} \& {Burkhart}}{{Krumholz} \&
  {Burkhart}}{2016}]{Krumholz16}
{Krumholz} M.~R.,  {Burkhart} B.,  2016, \mn@doi [\mnras]
  {10.1093/mnras/stw434}, \href
  {http://adsabs.harvard.edu/abs/2016MNRAS.458.1671K} {458, 1671}

\bibitem[\protect\citeauthoryear{{Krumholz} \& {Kruijssen}}{{Krumholz} \&
  {Kruijssen}}{2015}]{KK15}
{Krumholz} M.~R.,  {Kruijssen} J.~M.~D.,  2015, \mn@doi [\mnras]
  {10.1093/mnras/stv1670}, \href
  {http://adsabs.harvard.edu/abs/2015MNRAS.453..739K} {453, 739}

\bibitem[\protect\citeauthoryear{{Law}, {Steidel}, {Erb}, {Larkin}, {Pettini},
  {Shapley}  \& {Wright}}{{Law} et~al.}{2009}]{Law09}
{Law} D.~R.,  {Steidel} C.~C.,  {Erb} D.~K.,  {Larkin} J.~E.,  {Pettini} M.,
  {Shapley} A.~E.,   {Wright} S.~A.,  2009, \mn@doi [\apj]
  {10.1088/0004-637X/697/2/2057}, \href
  {http://adsabs.harvard.edu/abs/2009ApJ...697.2057L} {697, 2057}

\bibitem[\protect\citeauthoryear{{Lehnert}, {Nesvadba}, {Le Tiran}, {Di
  Matteo}, {van Driel}, {Douglas}, {Chemin}  \& {Bournaud}}{{Lehnert}
  et~al.}{2009}]{Lehnert09}
{Lehnert} M.~D.,  {Nesvadba} N.~P.~H.,  {Le Tiran} L.,  {Di Matteo} P.,  {van
  Driel} W.,  {Douglas} L.~S.,  {Chemin} L.,   {Bournaud} F.,  2009, \mn@doi
  [\apj] {10.1088/0004-637X/699/2/1660}, \href
  {http://adsabs.harvard.edu/abs/2009ApJ...699.1660L} {699, 1660}

\bibitem[\protect\citeauthoryear{{Lehnert}, {Le Tiran}, {Nesvadba}, {van
  Driel}, {Boulanger}  \& {Di Matteo}}{{Lehnert} et~al.}{2013}]{Lehnert13}
{Lehnert} M.~D.,  {Le Tiran} L.,  {Nesvadba} N.~P.~H.,  {van Driel} W.,
  {Boulanger} F.,   {Di Matteo} P.,  2013, \mn@doi [\aap]
  {10.1051/0004-6361/201220555}, \href
  {http://adsabs.harvard.edu/abs/2013A%26A...555A..72L} {555, A72}

\bibitem[\protect\citeauthoryear{{Leroy}, {Bolatto}, {Simon}  \&
  {Blitz}}{{Leroy} et~al.}{2005}]{Leroy05}
{Leroy} A.,  {Bolatto} A.~D.,  {Simon} J.~D.,   {Blitz} L.,  2005, \mn@doi
  [\apj] {10.1086/429578}, \href
  {http://adsabs.harvard.edu/abs/2005ApJ...625..763L} {625, 763}

\bibitem[\protect\citeauthoryear{{Mac Low} \& {Klessen}}{{Mac Low} \&
  {Klessen}}{2004}]{MacLow04}
{Mac Low} M.-M.,  {Klessen} R.~S.,  2004, \mn@doi [Reviews of Modern Physics]
  {10.1103/RevModPhys.76.125}, \href
  {http://adsabs.harvard.edu/abs/2004RvMP...76..125M} {76, 125}

\bibitem[\protect\citeauthoryear{{Madau} \& {Dickinson}}{{Madau} \&
  {Dickinson}}{2014}]{Madau14}
{Madau} P.,  {Dickinson} M.,  2014, \mn@doi [\araa]
  {10.1146/annurev-astro-081811-125615}, \href
  {http://adsabs.harvard.edu/abs/2014ARA%26A..52..415M} {52, 415}

\bibitem[\protect\citeauthoryear{{Nesvadba}, {Lehnert}, {Eisenhauer},
  {Gilbert}, {Tecza}  \& {Abuter}}{{Nesvadba} et~al.}{2006}]{Nesvadba06}
{Nesvadba} N.~P.~H.,  {Lehnert} M.~D.,  {Eisenhauer} F.,  {Gilbert} A.,
  {Tecza} M.,   {Abuter} R.,  2006, \mn@doi [\apj] {10.1086/507266}, \href
  {http://adsabs.harvard.edu/abs/2006ApJ...650..693N} {650, 693}

\bibitem[\protect\citeauthoryear{{Owers} et~al.,}{{Owers}
  et~al.}{2017}]{Owers17}
{Owers} M.~S.,  et~al., 2017, \mn@doi [\mnras] {10.1093/mnras/stx562}, \href
  {http://adsabs.harvard.edu/abs/2017MNRAS.468.1824O} {468, 1824}

\bibitem[\protect\citeauthoryear{{Padoan}, {Federrath}, {Chabrier}, {Evans},
  {Johnstone}, {J{\o}rgensen}, {McKee}  \& {Nordlund}}{{Padoan}
  et~al.}{2014}]{Padoan14}
{Padoan} P.,  {Federrath} C.,  {Chabrier} G.,  {Evans} II N.~J.,  {Johnstone}
  D.,  {J{\o}rgensen} J.~K.,  {McKee} C.~F.,   {Nordlund} {\AA}.,  2014,
  \mn@doi [Protostars and Planets VI]
  {10.2458/azu_uapress_9780816531240-ch004}, \href
  {http://adsabs.harvard.edu/abs/2014prpl.conf...77P} {pp 77--100}

\bibitem[\protect\citeauthoryear{{Robotham} et~al.,}{{Robotham}
  et~al.}{2014}]{Robotham14}
{Robotham} A.~S.~G.,  et~al., 2014, \mn@doi [\mnras] {10.1093/mnras/stu1604},
  \href {http://adsabs.harvard.edu/abs/2014MNRAS.444.3986R} {444, 3986}

\bibitem[\protect\citeauthoryear{{Roman-Duval}, {Jackson}, {Heyer}, {Rathborne}
   \& {Simon}}{{Roman-Duval} et~al.}{2010}]{Roman-Duval10}
{Roman-Duval} J.,  {Jackson} J.~M.,  {Heyer} M.,  {Rathborne} J.,   {Simon} R.,
   2010, \mn@doi [\apj] {10.1088/0004-637X/723/1/492}, \href
  {http://adsabs.harvard.edu/abs/2010ApJ...723..492R} {723, 492}

\bibitem[\protect\citeauthoryear{{S{\'a}nchez}}{{S{\'a}nchez}}{2015}]{Sanchez15}
{S{\'a}nchez} S.~F.,  2015, in {Ziegler} B.~L.,  {Combes} F.,  {Dannerbauer}
  H.,   {Verdugo} M.,  eds,  IAU Symposium Vol. 309, Galaxies in 3D across the
  Universe. pp 85--92 (\mn@eprint {arXiv} {1410.0295}),
  \mn@doi{10.1017/S1743921314009375}

\bibitem[\protect\citeauthoryear{{S{\'a}nchez} et~al.,}{{S{\'a}nchez}
  et~al.}{2012}]{Sanchez12}
{S{\'a}nchez} S.~F.,  et~al., 2012, \mn@doi [\aap]
  {10.1051/0004-6361/201117353}, \href
  {http://adsabs.harvard.edu/abs/2012A%26A...538A...8S} {538, A8}

\bibitem[\protect\citeauthoryear{{Scalo} \& {Elmegreen}}{{Scalo} \&
  {Elmegreen}}{2004}]{Scalo04}
{Scalo} J.,  {Elmegreen} B.~G.,  2004, \mn@doi [\araa]
  {10.1146/annurev.astro.42.120403.143327}, \href
  {http://adsabs.harvard.edu/abs/2004ARA%26A..42..275S} {42, 275}

\bibitem[\protect\citeauthoryear{{Sharp} et~al.,}{{Sharp}
  et~al.}{2006}]{Sharp06}
{Sharp} R.,  et~al., 2006, in Society of Photo-Optical Instrumentation
  Engineers (SPIE) Conference Series. p. 62690G (\mn@eprint {}
  {astro-ph/0606137}), \mn@doi{10.1117/12.671022}

\bibitem[\protect\citeauthoryear{{Sharp} et~al.,}{{Sharp}
  et~al.}{2015}]{Sharp15}
{Sharp} R.,  et~al., 2015, \mn@doi [\mnras] {10.1093/mnras/stu2055}, \href
  {http://adsabs.harvard.edu/abs/2015MNRAS.446.1551S} {446, 1551}

\bibitem[\protect\citeauthoryear{{Stilp}, {Dalcanton}, {Skillman}, {Warren},
  {Ott}  \& {Koribalski}}{{Stilp} et~al.}{2013}]{Stilp13}
{Stilp} A.~M.,  {Dalcanton} J.~J.,  {Skillman} E.,  {Warren} S.~R.,  {Ott} J.,
   {Koribalski} B.,  2013, \mn@doi [\apj] {10.1088/0004-637X/773/2/88}, \href
  {http://adsabs.harvard.edu/abs/2013ApJ...773...88S} {773, 88}

\bibitem[\protect\citeauthoryear{{Tacconi} et~al.,}{{Tacconi}
  et~al.}{2010}]{tacconi10}
{Tacconi} L.~J.,  et~al., 2010, \mn@doi [\nat] {10.1038/nature08773}, \href
  {http://adsabs.harvard.edu/abs/2010Natur.463..781T} {463, 781}

\bibitem[\protect\citeauthoryear{{Tamburro}, {Rix}, {Leroy}, {Mac Low},
  {Walter}, {Kennicutt}, {Brinks}  \& {de Blok}}{{Tamburro}
  et~al.}{2009}]{Tamburro09}
{Tamburro} D.,  {Rix} H.-W.,  {Leroy} A.~K.,  {Mac Low} M.-M.,  {Walter} F.,
  {Kennicutt} R.~C.,  {Brinks} E.,   {de Blok} W.~J.~G.,  2009, \mn@doi [\aj]
  {10.1088/0004-6256/137/5/4424}, \href
  {http://adsabs.harvard.edu/abs/2009AJ....137.4424T} {137, 4424}

\bibitem[\protect\citeauthoryear{{Tasker} \& {Tan}}{{Tasker} \&
  {Tan}}{2009}]{Tasker09}
{Tasker} E.~J.,  {Tan} J.~C.,  2009, \mn@doi [\apj]
  {10.1088/0004-637X/700/1/358}, \href
  {http://adsabs.harvard.edu/abs/2009ApJ...700..358T} {700, 358}

\bibitem[\protect\citeauthoryear{{Taylor} et~al.,}{{Taylor}
  et~al.}{2011}]{Taylor11}
{Taylor} E.~N.,  et~al., 2011, \mn@doi [\mnras]
  {10.1111/j.1365-2966.2011.19536.x}, \href
  {http://adsabs.harvard.edu/abs/2011MNRAS.418.1587T} {418, 1587}

\bibitem[\protect\citeauthoryear{{Varidel}, {Pracy}, {Croom}, {Owers}  \&
  {Sadler}}{{Varidel} et~al.}{2016}]{Varidel16}
{Varidel} M.,  {Pracy} M.,  {Croom} S.,  {Owers} M.~S.,   {Sadler} E.,  2016,
  \mn@doi [\pasa] {10.1017/pasa.2016.3}, \href
  {http://adsabs.harvard.edu/abs/2016PASA...33....6V} {33, e006}

\bibitem[\protect\citeauthoryear{{Veilleux} \& {Osterbrock}}{{Veilleux} \&
  {Osterbrock}}{1987}]{Veilleux87}
{Veilleux} S.,  {Osterbrock} D.~E.,  1987, \mn@doi [\apjs] {10.1086/191166},
  \href {http://adsabs.harvard.edu/abs/1987ApJS...63..295V} {63, 295}

\bibitem[\protect\citeauthoryear{{Wilson} et~al.,}{{Wilson}
  et~al.}{2011}]{Wilson11}
{Wilson} C.~D.,  et~al., 2011, \mn@doi [\mnras]
  {10.1111/j.1365-2966.2010.17646.x}, \href
  {http://adsabs.harvard.edu/abs/2011MNRAS.410.1409W} {410, 1409}

\bibitem[\protect\citeauthoryear{{Wisnioski} et~al.,}{{Wisnioski}
  et~al.}{2015}]{Wisnioski15}
{Wisnioski} E.,  et~al., 2015, \mn@doi [\apj] {10.1088/0004-637X/799/2/209},
  \href {http://adsabs.harvard.edu/abs/2015ApJ...799..209W} {799, 209}

\bibitem[\protect\citeauthoryear{{van de Sande} et~al.,}{{van de Sande}
  et~al.}{2017}]{Sande17}
{van de Sande} J.,  et~al., 2017, \mn@doi [\apj] {10.3847/1538-4357/835/1/104},
  \href {http://adsabs.harvard.edu/abs/2017ApJ...835..104V} {835, 104}

\makeatother
\end{thebibliography}

%%%%%%%%%%%%%%%%%%%%%%%%%%%%%%%%%%%%%%%%%%%%%%%%%%

%%%%%%%%%%%%%%%%% APPENDICES %%%%%%%%%%%%%%%%%%%%%

%%%%%%%%%%%%%%%%%%%%%%%%%%%%%%%%%%%%%%%%%%%%%%%%%%%

% Don't change these lines
\bsp	% typesetting comment
\label{lastpage}

\end{document}